\documentclass[aps,prb,reprint,showpacs,superscriptaddress]{revtex4-1}
\pdfoutput=1
\usepackage{graphicx}
\usepackage{amsmath}
\usepackage{amsfonts}
\usepackage{color}
\usepackage{hyperref}
\usepackage{multirow}
\usepackage[usenames,dvipsnames]{xcolor}
\usepackage{subfigure}
\usepackage{epstopdf}

\newcommand{\Ket}[1]{| #1 \rangle}

\newcommand{\Braket}[1]{\langle #1 \rangle}

\newcommand{\MPQ}{Max-Planck-Institut f\"ur Quantenoptik, Hans-Kopfermann-Str.\ 1, D-85748 Garching, Germany}
\newcommand{\madrid}{Instituto de F\'isica Te\'orica, UAM-CSIC, Madrid, Spain}

\newcommand{\dresden}{Max-Planck-Institut f\"ur Physik komplexer Systeme, D-01187 Dresden, Germany}
\newcommand{\aarhus}{Department of Physics and Astronomy, Aarhus University, DK-8000 Aarhus C,
Denmark}

\makeatletter
\newcommand{\redboxed}[1]{\textcolor{red}{%
  \fbox{\normalcolor\m@th$\displaystyle#1$}}}
\makeatother

\begin{document}


\title{Lattice effects on Laughlin wave functions\\ and parent Hamiltonians}

\begin{abstract}
We investigate lattice effects on wave functions that are lattice analogues of bosonic and fermionic Laughlin wave functions with number of particles per flux $\nu=1/q$ in the Landau levels. These wave functions are defined analytically on lattices with $\mu$ particles per lattice site, where $\mu$ may be different than $\nu$. We give numerical evidence that these states have the same topological properties as the corresponding continuum Laughlin states for different values of $q$ and for different fillings $\mu$. These states define, in particular, particle-hole symmetric lattice Fractional Quantum Hall states when the lattice is half-filled. On the square lattice it is observed that for $q\leq 4$ this particle-hole symmetric state displays the topological properties of the continuum Laughlin state at filling fraction $\nu=1/q$, while for larger $q$ there is a transition towards long-range ordered anti-ferromagnets. This effect does not persist if the lattice is deformed from a square to a triangular lattice, or on the Kagome lattice, in which case the topological properties of the state are recovered. We then show that changing the number of particles while keeping the expression of these wave functions identical gives rise to edge states that have the same correlations in the bulk as the reference lattice Laughlin states but a different density at the edge. We derive an exact parent Hamiltonian for which all these edge states are ground states with different number of particles. In addition this Hamiltonian admits the reference lattice Laughlin state as its unique ground state of filling factor $1/q$. Parent Hamiltonians are also derived for the lattice Laughlin states at other fillings of the lattice, when $\mu\leq 1/q$ or $\mu\geq 1-1/q$ and when $q=4$ also at half-filling.
\end{abstract}


\author{Ivan Glasser}
\affiliation{\MPQ}
\author{J. Ignacio Cirac}
\affiliation{\MPQ}
\author{Germ\'an Sierra}
\affiliation{\madrid}
\author{Anne E. B. Nielsen}
\affiliation{\MPQ}\affiliation{\aarhus}\affiliation{\dresden}
\maketitle

\section{Introduction}
\label{SEC:Introduction}

Strongly correlated quantum systems can give rise to fascinating phenomena that are absent from conventional materials. For example, topological states of matter have attracted a lot of attention, both for the fundamental physics they display as well as for their potential practical applications. They were first realized experimentally with the discovery of the Fractional Quantum Hall (FQH) effect\cite{Tsui1982}, in which electrons of a two-dimensional electron gas subject to a strong magnetic field form an incompressible quantum fluid giving rise to fractionally charged excitations. A large understanding of the FQH effect was made possible by the discovery of analytical wave functions describing the electrons in a partially filled Landau level, such as Laughlin's wave function\cite{Laughlin1983}.

Experimental realizations of FQH states and manipulation of their quasiparticles however remain a challenge, which motivated a large body of research devoted to the realization of the FQH effect in different systems. An example of such possible systems are chiral spin liquids. In this phase of matter, spins on a lattice form a collective state that has the same topological properties as FQH states. They were first introduced with the Kalmeyer-Laughlin state\cite{Kalmeyer1987}, a bosonic Laughlin state at filling fraction $\nu=1/2$ on a lattice with emergent anyonic excitations. Over the last decades, much research has been devoted to finding spin Hamiltonians that have such a state as their ground state. Parent Hamiltonians have been obtained for the Kalmeyer-Laughlin state and its generalizations\cite{Schroeter2007,Thomale2009,Kapit2010,Nielsen2012,Greiter2014,Tu2014} and recently several local Hamiltonians on the square or Kagome lattice have been shown to exhibit such a chiral spin liquid as their ground state \cite{Nielsen2013,Bauer2014,Gong2014,He2014,Gong2015,He2015,Wietek2014,Hu2015,Kumar2015bis,Hu2016}. Parent Hamiltonians and local Hamiltonians have been found as well for a more exotic non-Abelian chiral spin liquid corresponding to the Moore-Read state \cite{Greiter2009,Greiter2014,Glasser2015}, while another approach has been the realization of these states in topological flat-band models\cite{Sheng2011,Neupert2011,Wang2011,Sun2011,Regnault2011}.

Typically, the previous examples of the realization of the Kalmeyer-Laughlin state on spin-$1/2$ lattices occur either at half-filling of the lattice (half of the spins are pointing up in the $z$-direction) or at low filling of the lattice (so the system is close to a continuum limit), but it is also proposed that this state can emerge on a lattice at filling $1/3$ \cite{Kumar2014}. Indeed it was recently observed\cite{Zhu2015} numerically that a bosonic FQH Laughlin state with number of particles per flux $\nu=1/2$ appears as ground state of a Bose-Hubbard  model on the Kagome lattice in the hard-core limit when the boson filling fraction is $1/3$. Moreover, exact parent Hamiltonians have been obtained for lattice versions of the Laughlin states at number of particles per flux $\nu=1/q$ ($q\in  \mathbb{N}$) only when the filling factor $\mu$ of the lattice is equal to $\nu$, although it is possible to define the Laughlin states also on lattices with a different filling factor\cite{Tu2014}. This raises the question of when Laughlin states with number of particles per flux $\nu=1/q$ can appear on lattices when the filling of the lattice $\mu$ is different than $\nu$. This question is relevant for experimental purposes: indeed, if Hamiltonians can be found to stabilize these states when the lattice filling factor is not the same as the continuum filling fraction, this means that there is more flexibility in realizing different models displaying FQH physics in new settings.

It is also of interest to understand whether the lattice may destroy the physics of the FQH effect. Indeed, although one way to check whether a lattice Hamiltonian reproduces FQH physics has been to compare overlaps between its ground state and FQH wave functions\cite{Hafezi2007,Nielsen2013}, it is not a priori clear that a wave function of a FQH state in the continuum still displays the same topological effects once discretized and placed on a lattice. It has been shown to be the case for the Kalmeyer-Laughlin wave function\cite{Zhang2011,Zhang2012,Nielsen2012} as well as for the Moore-Read wave function\cite{Glasser2015,Wildeboer2015}, but we will show here that while it remains true for most of the wave functions we introduce, in some particular cases strong lattice effects may destroy the topological properties.

In this work we address these issues by studying lattice versions of Laughlin wave functions at $\nu=1/q$ for bosons and fermions on arbitrary lattices with filling factor $\mu$ not necessarily equal to $\nu$, thus generalizing the Kalmeyer-Laughlin state in the lines of Ref.~\onlinecite{Tu2014}. We investigate the phase diagram for different values of $\mu$ and $q$ using measures of correlations, entanglement entropy and braiding of localized quasiholes to give numerical evidence that in a large part of the phase diagram the states on a square lattice have the same topological properties as the continuum Laughlin states. 

When the lattice is half-filled the states have the additional property that they are symmetric under particle-hole transformation. While the states for $q\leq 4$ define lattice versions of Laughlin states at $\nu=1/q$, even on a half-filled lattice, it is found that at half-filling on the square lattice the states for $q\geq 5$ display long-range anti-ferromagnetic order and are topologically trivial. It is shown that this behaviour is due to geometric properties of the square lattice and that it disappears on frustrated lattices like the triangular and the Kagome lattice, so that states having the topological properties of the Laughlin state at $\nu=1/q$, $q\geq 2$, can be defined on such half-filled frustrated lattice.

We define edge state wave functions for the lattice Laughlin states by acting on the state with a charge located outside the boundary of the lattice. This provides a second way to change the filling of the lattice without changing the topological properties of the state and we show that the edge states that are obtained in this way have a density which is modified at the edge compared to the original lattice Laughlin states, while their density and correlations in the bulk are the same. The states thus constructed are different from the edge states for the Kalmeyer-Laughlin state introduced in Ref.~\onlinecite{Herwerth2015}, but they share similar properties.

In addition we derive an exact parent Hamiltonian for which all these edge states are ground states with different number of particles. This Hamiltonian also admits the corresponding lattice Laughlin state as its unique ground state of filling factor $1/q$. Parent Hamiltonians are also derived for the lattice Laughlin states at other fillings of the lattice, when $\mu\leq 1/q$ or $\mu\geq 1-1/q$ and when $q=4$ also at half-filling, which significantly extends the group of models for which parent Hamiltonians can be found. It is notable that these Hamiltonians are not more complicated than the previous ones, but simply require different coupling strengths. The wave functions we consider are invariant under conformal transformations of the lattice positions, but the Hamiltonians we find do not display this symmetry. This allows us to construct a more general class of Hamiltonians that has the lattice Laughlin states as their ground state. Among this class it is possible to find a Hamiltonian which is invariant under the symmetry transformations of the lattice on the cylinder, but there is no Hamiltonian in this class that would be invariant under all conformal transformations. These parent Hamiltonians are long-ranged and involve up to three-body or up to four-body interactions, depending on the model considered. 

The paper is organized as follows : in Section \ref{section1}, we define lattice Laughlin states at different lattice filling factors. In Section \ref{sectionprop}, we compute their properties on the square lattice. In Section \ref{section3} we discuss the particular case of half-filling and show that for large $q$ the states display topological order on frustrated lattices while they are trivial anti-ferromagnets on the square lattice. In Section \ref{sectionedge} we introduce and characterize edge states for the lattice Laughlin states and in Section \ref{sectionparentH} we derive the exact parent Hamiltonians.

\section{Lattice Laughlin States at different lattice filling factors}

\label{section1}
In this section we define Laughlin wave functions with number of particles per flux $\nu=1/q$ ($q\in  \mathbb{N}$) on lattices with arbitrary lattice filling factor, defined as the ratio of the number of particles and the number of lattice sites.

Let us consider an integer $N$ and a lattice with $N$ sites at positions $z_j$, $j\in\{1, 2, \ldots , N\}$ in the complex plane. Particular choices of the coordinates $z_j$ will for example yield a square, triangular or Kagome lattice on the plane or on the cylinder (using a mapping of the coordinates from the cylinder to the plane). At each lattice site $z_i$ we associate a local basis $\Ket{n_i}$ such that the lattice site can be empty ($n_i=0$) or occupied ($n_i=1$). We consider $M$ particles (hardcore bosons or fermions) hopping on this lattice and define $\mu\equiv\frac{M}{N}$ as the lattice filling factor. 

In general, a wave function defined on this lattice will have the form
\begin{align}
\Ket{\psi} &= \sum_{n_1, \dots, n_N} \psi(n_1, \ldots, n_N) \Ket{n_1, \dots, n_N},
\end{align}
where $\psi(n_1, \ldots, n_N)=0$ unless the number of particles is $\sum_{i} n_i=M$ (if the particles are fermions, the Fock states are defined using the same ordering as for the lattice sites).
Let $q$ be a positive integer. To define Laughlin wave functions at $\nu=1/q$ on this lattice, we follow the idea of Moore and Read \cite{Moore1991}, later extended to lattices \cite{cftimps,Nielsen2012,Tu2014}, to express the wave function as some correlator of operators from Conformal Field Theory (CFT):
\begin{align}
\label{correlator}
\psi(n_1, \ldots, n_N)&\propto \Braket{V_{n_1}(z_1) \dots V_{n_N}(z_N)},
\end{align}
where $V_{n_i}(z_i)$ are operators attached at position $z_j$.

At each lattice site we attach the vertex operators
\begin{align}
\label{voperators}
V_{n_j}(z_j)=
\begin{cases}
    :e^{-i\frac{\eta}{\sqrt{q}}\phi (z_{j})}: & \text{if $n_j=0$},\\
    e^{i\pi\eta'(j-1)}:e^{i\frac{q-\eta}{\sqrt{q}} \phi (z_{j})}: & \text{if $n_j=1$},
  \end{cases}
\end{align}
where $\phi(z)$ is a chiral bosonic field from the $c=1$ Conformal Field Theory, $:\ldots:$ denotes normal ordering and $\eta$ is a positive rational number such that $\eta N/q$ is an integer and in all this work $\eta'$ denotes $\min(\eta,q-\eta)$. If $\eta$ is not an integer, a choice of branch cuts can be made consistently for all the formulas in this work.

Evaluating the correlator in Eq.~\eqref{correlator} yields\cite{cft,Tu2014} a wave function, that we denote as $\psi_q^\eta$, such that
\begin{align}
\label{Laughlinstate}
\psi_q^\eta(n_1, \ldots, n_N)\propto &\delta_n \xi_{\eta} \prod_{i<k}(z_i-z_k)^{q n_i n_k} \prod_{j\neq l} (z_{l}-z_{j})^{-\eta n_l},
\end{align}
where $\delta_n$ is zero unless the number of particles is $M=\sum_i n_i=\eta\frac{N}{q}$ and $\xi_\eta$ is $1$ if $\eta\leq q-\eta$ or $(-1)^{q\sum_j (j-1)n_j}$ otherwise. The first factor in Eq.~\eqref{Laughlinstate} can be interpreted as the attachement of a flux $q$ to each particle, while the second factor represents the net effect of the lattice on a given particle, that in the continuum is generated by the background charge. Note that no background charge needs to be inserted to evaluate the correlator and charge neutrality is directly ensured by the choice of operators \eqref{voperators}. 

It was shown in Ref.~\onlinecite{Nielsen2012,Tu2014} for lattices defined on the plane that in the thermodynamic limit and when the area $a$ of each site is the same, 
\begin{align}
\prod_{j\neq l} (z_{l}-z_{j})^{-\eta n_l} \propto e^{-i\sum_l g_l} e^{-\frac{1}{4}\frac{2\pi\eta}{a}\sum_l |z_l|^2 n_l},
\end{align}
where $g_l\equiv \text{Im} [\eta \sum_{j(\neq l)}n_l\ln(z_l-z_j)]$ is a real number that gives rise to a phase factor that does not change the correlations or entanglement entropy of the state. Moreover it was shown that the approximation is valid already for moderate sizes\cite{Nielsen2012,Tu2014}. In the continuum, the wave functions are usually expressed in the basis spanned by the position of the particles $Z_i$. Here the $Z_i$ are the positions $z_i$ where $n_i=1$ and the $(Z_1,...,Z_M)$ form a basis of the Hilbert space. The wave function written in this basis then becomes (here the phase factors, which can be transformed away if desired, are omitted)
\begin{align}
\Psi_\text{Laughlin}(Z_1, \ldots, Z_M)\propto & \prod_{i<j}(Z_i-Z_j)^{q} e^{-\frac{1}{4}\frac{2\pi\eta}{a}\sum_l |Z_l|^2},
\end{align}
which is the Laughlin wave function at filling fraction $\nu=\frac{1}{q}$ for $M$ particles with positions restricted to the lattice sites. 

To define the states on the cylinder, the coordinates are taken to be of the form $z_j=e^{x_j+iy_j}$, where $x_j+iy_j$ define the corresponding lattice on the plane. The previous factor can then be computed by writing 
\begin{align}
\prod_{j\neq l} (z_{l}-z_{j})^{-\eta n_l} \propto e^{-\sum_l n_l \sum_{j(\neq l)} \eta \log(z_{l}-z_{j})},
\end{align}
and in the thermodynamic limit the real part of the sum can be replaced by an integral\cite{Tu2014,Moore1991} on the cylinder :
\begin{align}
\sum_{j(\neq l)} \eta \log(|z_{l}-z_{j}|)\propto \frac{\eta}{a}\int_{x=0}^{R} \int_{y=0}^{2\pi}  \log(|z_{l}-e^x e^{iy}|) \text{d}y \text{d}x ,
\end{align}
which can be evaluated to yield $-\frac{2\pi \eta}{a}\frac{|x_l|^2}{2}+\text{constant}$, where the constant does not depend on $z_l$. In the thermodynamic limit on the cylinder the wave function therefore also reduces to the Laughlin wave function with positions restricted to the lattice sites.

Here it is important to note that the filling factor of the lattice is $\mu=\frac{M}{N}=\frac{\eta}{q}$, which can be different from the Laughlin filling fraction $\nu=1/q$. Here $\mu$ is the number of particles per lattice site, whereas $\nu$ is the number of particles per number of fluxes. In this work we will be particularly interested in lattice effects in the cases where $\eta\neq 1$, so that $\mu\neq \nu$.

\begin{figure}[htb]
\includegraphics[scale=0.55]{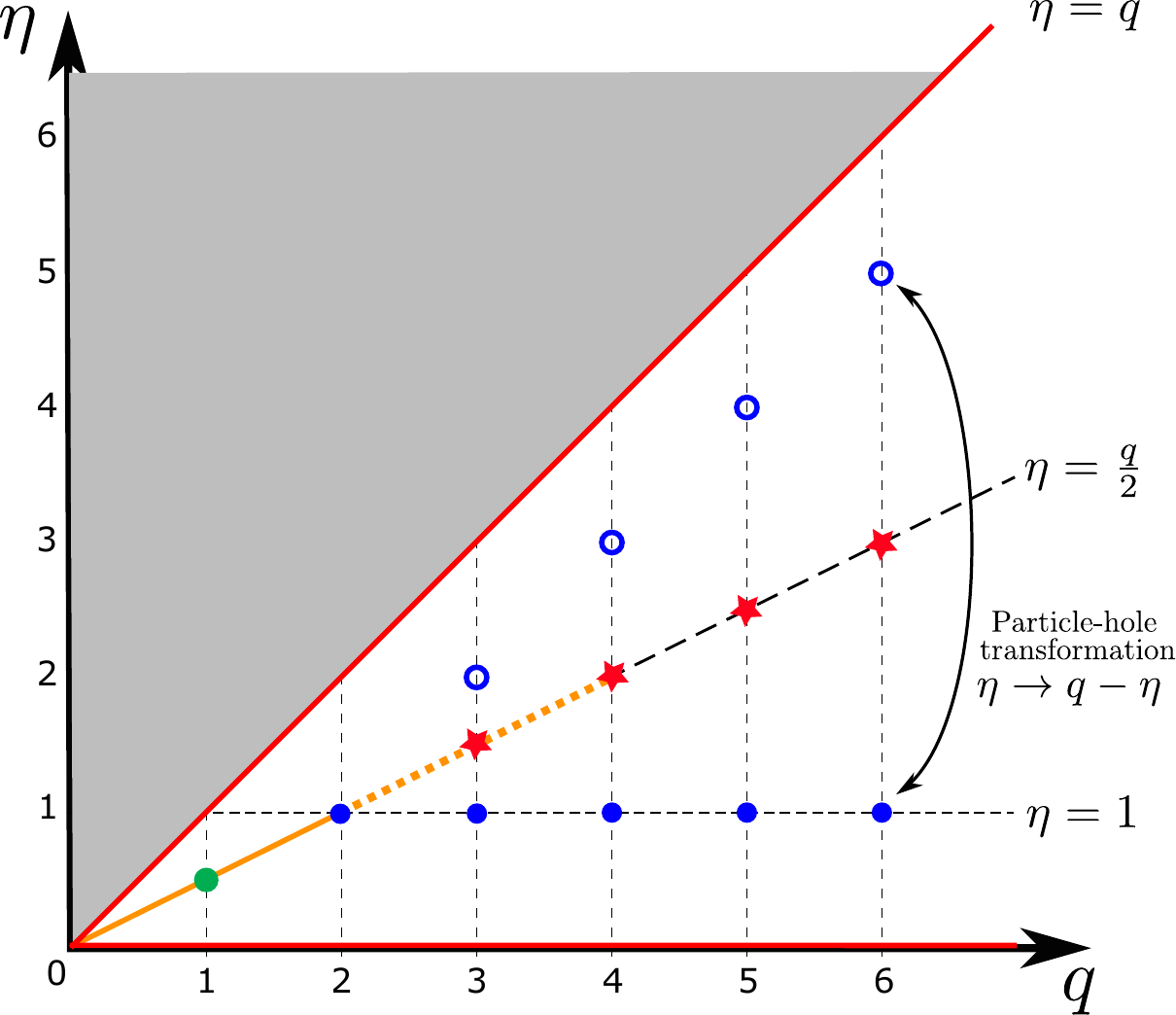}
\caption{\label{schemadebase}
  Diagram of the $\psi_q^\eta$ states for the values of $q$ and $\eta$ considered in this work. $q$ is integer while $\eta$ is a rational number in the interval $[0,q]$, since the states for $\eta>q$ (grey region) are not defined. When $\eta\rightarrow 0$ or $\eta\rightarrow q$ (red lines), the state is a ferromagnet if $N$ is fixed, or a Laughlin state in the continuum if $N$ goes to infinity and $N\propto1/\eta$ (if $\eta\rightarrow 0$) or $N\propto1/(q-\eta)$ (if $\eta\rightarrow q$) . The solid blue disks on the line $\eta=1$ correspond to lattice Laughlin states that satisfy $\mu=\nu=1/q$. The green disk is an Integer Quantum Hall state. On the orange line (q not necessarily integer), the state in 1D is critical, while on the dotted orange line it has long range order.
}
\end{figure}

Let us first discuss a few particular values of the parameters $q$ and $\eta$. In Fig.~\ref{schemadebase}, we draw a diagram of the $\psi_q^\eta$ states for the different values of the parameters that will be considered in this work. ~
When $\mu=0$ (resp. $\mu=1$), all the lattice sites are empty (resp. occupied), so the state is a trivial ferromagnet. This happens when $\eta=0$ or when $\eta=q$ (red lines in Fig.~\ref{schemadebase}). When $\eta>q$, the number of particles in the lattice has to be larger than the number of lattice sites, which is not possible since only single occupancy is allowed. The state $\psi_q^\eta$ is therefore not defined for $\eta>q$ (grey region in Fig.~\ref{schemadebase}).

Particular cases that have been considered before are the $\eta=1$, $q\geq 2$ states (blue disks in Fig.~\ref{schemadebase}), for which the states have been shown\cite{Tu2014} to be lattice versions of the Laughlin states and parent Hamiltonians have been obtained. When $\eta\leq 1$, it was shown\cite{Tu2014} that the topological entanglement entropy of the states remains the same and that in the limit $\eta\rightarrow 0$, $N\propto 1/\eta \rightarrow \infty$, the states become Laughlin states in the continuum.

 Another particularly interesting case is the choice $\eta=q/2$, which corresponds to half-filling of the lattice. In one dimension, the states defined at half-filling (and q not necessarily integer) have been shown\cite{cftimps} to be critical states for $q\leq 2$ (orange line in Fig.~\ref{schemadebase}) and to exhibit antiferromagnetic long-range order for $2<q<4$ (dotted orange line in Fig.~\ref{schemadebase}). At $q=1$, $\eta=1/2$ (green disk), the state in 2D corresponds to an Integer Quantum Hall state\cite{cftimps}. Note that the states are bosonic for even $q$ and fermionic for odd $q$.

\section{Properties of the Lattice Laughlin States}
\label{sectionprop}

In this section we explore the properties of the lattice Laughlin States to determine the phase diagram of the states in the $(q,\eta)$ plane. Particle-hole transformation shows that the phase diagram is symmetric along the $\eta=q/2$ line, on which the states are particle-hole symmetric. We then give numerical evidence that the states have exponentially decaying correlations on the square lattice unless $\eta=q/2$, $q\geq 5$. Moreover we show that when the correlations decay exponentially, the states have the same topological entanglement entropy as the Laughlin states in the continuum and that quasihole excitations of these states have the corresponding anyonic statistics.

\subsection{Particle-hole transformation}

The different wave functions $\psi_q^\eta$ are related under particle-hole transformation. This transformation has the effect of exchanging $n_i$ and $1-n_i$, so the vertex operators used in the definition of the wave function are transformed as
\begin{align}
    V_{n_j}(z_j)&=e^{i\pi \eta' (j-1)n_j}:e^{i\frac{qn_j-\eta}{\sqrt{q}}\phi (z_{j})}:\longrightarrow V_{n_j}'(z_j),
\end{align}
where 
\begin{align}
    V_{n_j}'(z_j)&\equiv e^{i\pi \eta' (j-1)}e^{-i\pi \eta' (j-1)n_j}:e^{-i\frac{qn_j-(q-\eta)}{\sqrt{q}}\phi (z_{j})}:.
\end{align}
Now let us define $\tilde{V}_{n_i}(z_j)$ the operators used to define the state $\psi_q^{q-\eta}$ and observe that
\begin{align}
\Braket{V_{n_1}'(z_1)  \dots V_{n_N}'(z_N)}\propto\Braket{\tilde{V}_{n_1}(z_1) \dots \tilde{V}_{n_N}(z_N)}.
\end{align}
This shows that the states $\psi_q^\eta$ and $\psi_q^{q-\eta}$ are exchanged under particle-hole transformation (see Fig.~\ref{schemadebase}). Note that this transformation also changes the number of particles, so that it relates states at lattice filling factor $1/q$ and $1-1/q$. This can be compared to the situation in the continuum, where a particle-hole transformation can be defined\cite{Girvin1984} to relate the Fractional Quantum Hall states at filling fraction $1/q$ and $1-1/q$. On the lattice the particle-hole transformation is however different, because it only involves exchanging the $\Ket{0}$ and $\Ket{1}$ states, so there is no separate treatment of an electron or a hole. The $\psi_q^\eta$ and the $\psi_q^{q-\eta}$ states are therefore related by a simple change of basis, and as such all properties of the wave function that are symmetric with respect to particle-hole transformation will be the same for the $\psi_q^\eta$ and the $\psi_q^{q-\eta}$ states. This is in particular the case of the connected particle-particle correlation function, since $\Braket{n_i n_j}-\Braket{n_i}\Braket{n_j}=\Braket{(1-n_i) (1-n_j)}-\Braket{1-n_i}\Braket{1-n_j}$.

\begin{figure}[htb]
\includegraphics[scale=0.35]{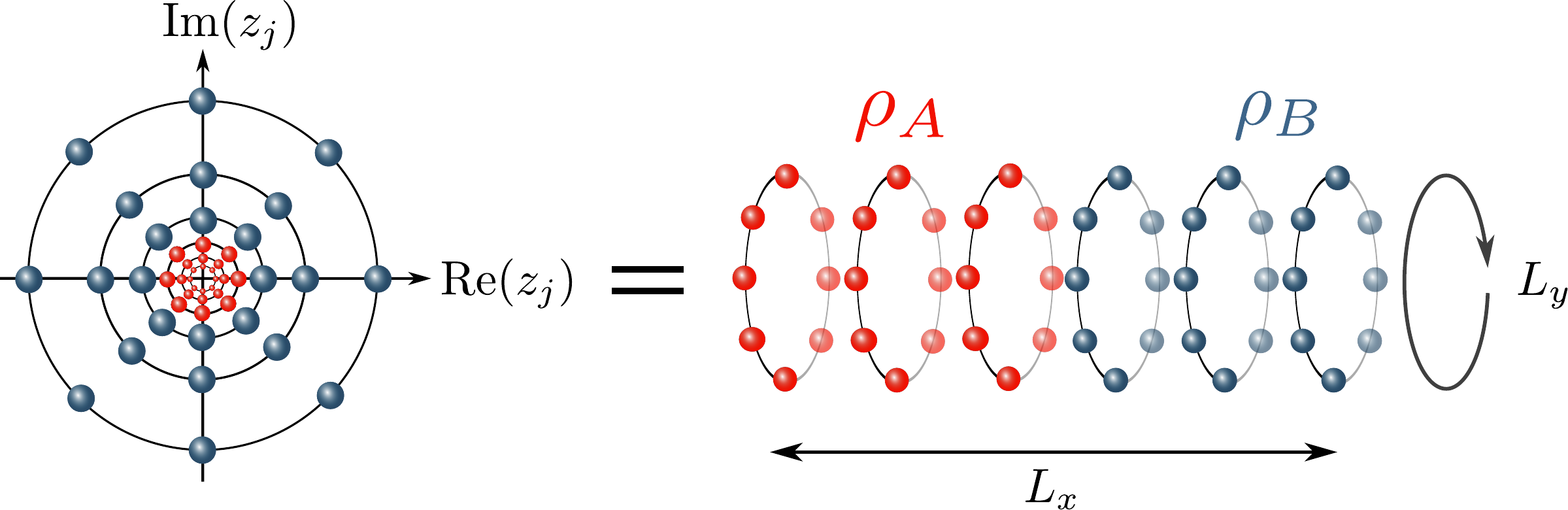}
\caption{\label{fig:cylinder}
  The mapping from a lattice on the complex plane to a cylinder. For the square lattice on the cylinder, the coordinates are $z_j=\exp(2\pi((x_j-L_x/2+1/2)/L_y+(y_j-L_y/2+1/2)i/L_y))$, $x_j\in\{0,\ldots,L_x-1\}$, $y_j\in\{0,\ldots,L_y-1\}$. To compute the entanglement entropy of the state, the cylinder is cut into two halves and the Renyi entropy of the first half is computed using a Metropolis-Hastings algorithm. The topological entanglement entropy is extracted by varying the size $L_y$ of the cylinder.
}
\end{figure}

\begin{figure}[htb]
\includegraphics[scale=0.52]{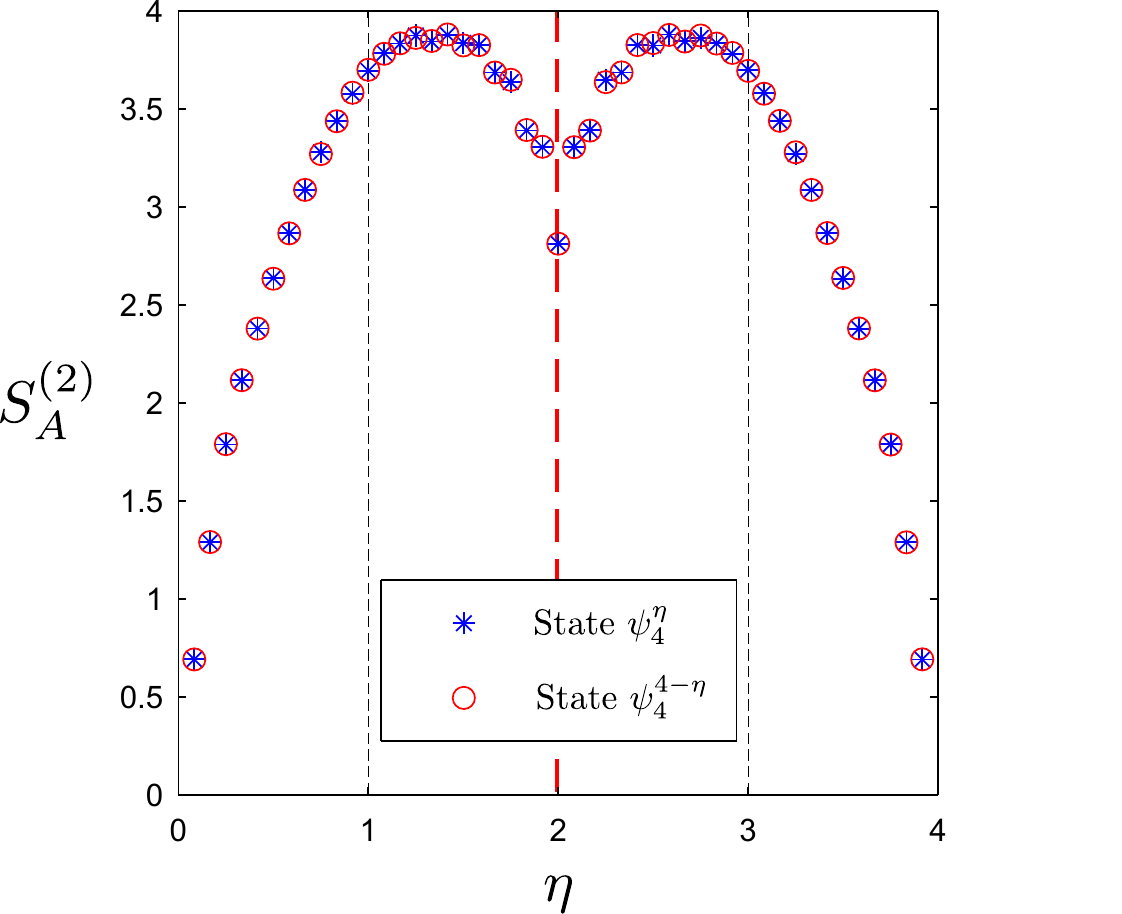}
\caption{\label{fig:entropysymm}
  Renyi entropy $S^{(2)}_A$ of the first half of the cylinder for the states $\psi_4^{\eta}$ and $\psi_4^{4-\eta}$ on a square $12\times 12$ lattice. The symmetry axis is shown as a red dashed line. The states at $\eta=1$ and $\eta=3$ have lattice filling factors $1/4$ and $3/4$ and correspond to the lattice versions of the $\nu=1/4$ and $\nu=3/4$ Laughlin states. The state at $\eta=2$ is particle-hole symmetric on a half-filled lattice. The errors estimated from the variation to the mean from Monte Carlo simulations with different initial conditions are below $0.01$ for all points.
}
\end{figure}

We further illustrate this symmetry by computing the Renyi entropy $S^{(2)}_A = - \ln {\rm Tr} \;  \rho^2_A$, where $\rho_A$ is the density matrix of the $\psi_q^\eta$ state restricted to the first half of the square lattice on a cylinder (see Fig.~\ref{fig:cylinder}). Since this quantity is invariant under a change of basis $\Ket{0}\leftrightarrow\Ket{1}$, this quantity is the same for the states $\psi_q^\eta$ and $\psi_q^{q-\eta}$. As an illustration we perform this computation by using a Metropolis-Hastings Monte Carlo algorithm with two independent spin chains \cite{Bajdich2008,cftimps,Hastings2010} for $q=4$ as shown in Fig.~\ref{fig:entropysymm}.

The states at $\eta=q/2$ are special with respect to this transformation, since these states are particle-hole symmetric : the operators used to define these states are $:e^{\pm i\sqrt{q}/2}:$. This is in particular the case for the bosonic $\psi_2^1$ state, as is the case for the bosonic $\nu=1/2$ Laughlin state in the continuum. For larger values of $q$, the Laughlin states in the continuum are not particle-hole symmetric. However on the lattice we can change independently the lattice filling factor $\mu$ to $1/2$, which amounts to considering the states $\psi_q^{q/2}$ (red stars in Fig.~\ref{schemadebase}). Fig.~\ref{fig:entropysymm} shows that these state have lower entanglement entropy than the states at $\eta=1$. The properties of these states will be investigated in more details in the following.

\subsection{Correlations in the phase diagram}

\begin{figure}[htb]
\includegraphics[scale=0.55]{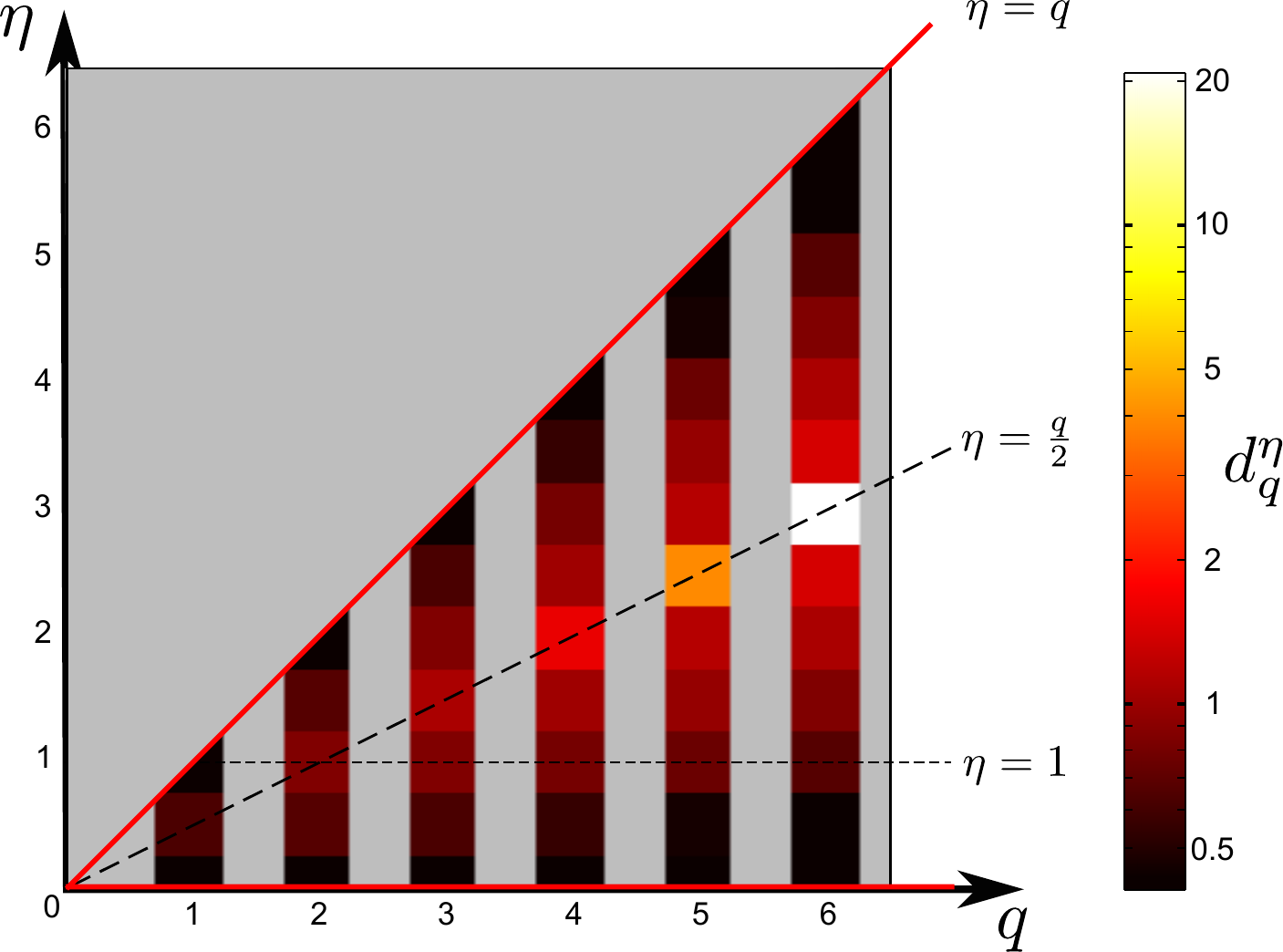}
\caption{\label{phasediagramcorrelation}
  Characteristic length $d_q^\eta$ (Eq.~\eqref{charlength}) estimating the correlation length of the $\psi_q^\eta$ states. The characteristic length is computed for $q$ integer and $\eta$ integer and half-integer using a Metropolis-Hastings algorithm on a square $12\times 12$ (or $10\times 12$ for $q=5$) lattice on the cylinder. The decay is found to be exponential everywhere, except when $\eta=q/2$, $q\geq 5$.
}
\end{figure}

To investigate possible different phases in the $(q,\eta)$-diagram, we start by computing the connected particle-particle correlation function $C_{ij}\equiv\Braket{n_i n_j}-\Braket{n_i}\Braket{n_j}$ in the bulk of the system for a square lattice on the cylinder, using a Metropolis-Hastings Monte Carlo algorithm. It is found that the correlation function decays exponentially with the distance for all values of $q$ and $\eta$, except for $q\geq 5$ at $\eta=q/2$. In Fig.~\ref{phasediagramcorrelation}, we explore the phase diagram by using as parameter the correlation length estimated from nearest-neighbours and next-nearest neighbours correlations in the bulk of the system. More precisely we compute a characteristic length, defined as 
\begin{align}
\label{charlength}
d_q^{\eta}=\frac{1}{\ln(C_{i(i+1)})-\ln(C_{i(i+2)})},
\end{align}
where $i$, $i+1$ and $i+2$ index neighbouring lattice sites in the periodic direction that are located on a ring in the middle of the cylinder. This parameter is an estimate on how fast the correlations decay. When the decay is exponential with a short correlation length, we observe numerically that this quantity almost coincides with the correlation length. When this characteristic length is larger, further investigations are performed to characterize the decay of correlations.

\begin{figure}[htb]
\subfigure[]{
        \centering
        \includegraphics[scale=0.58]{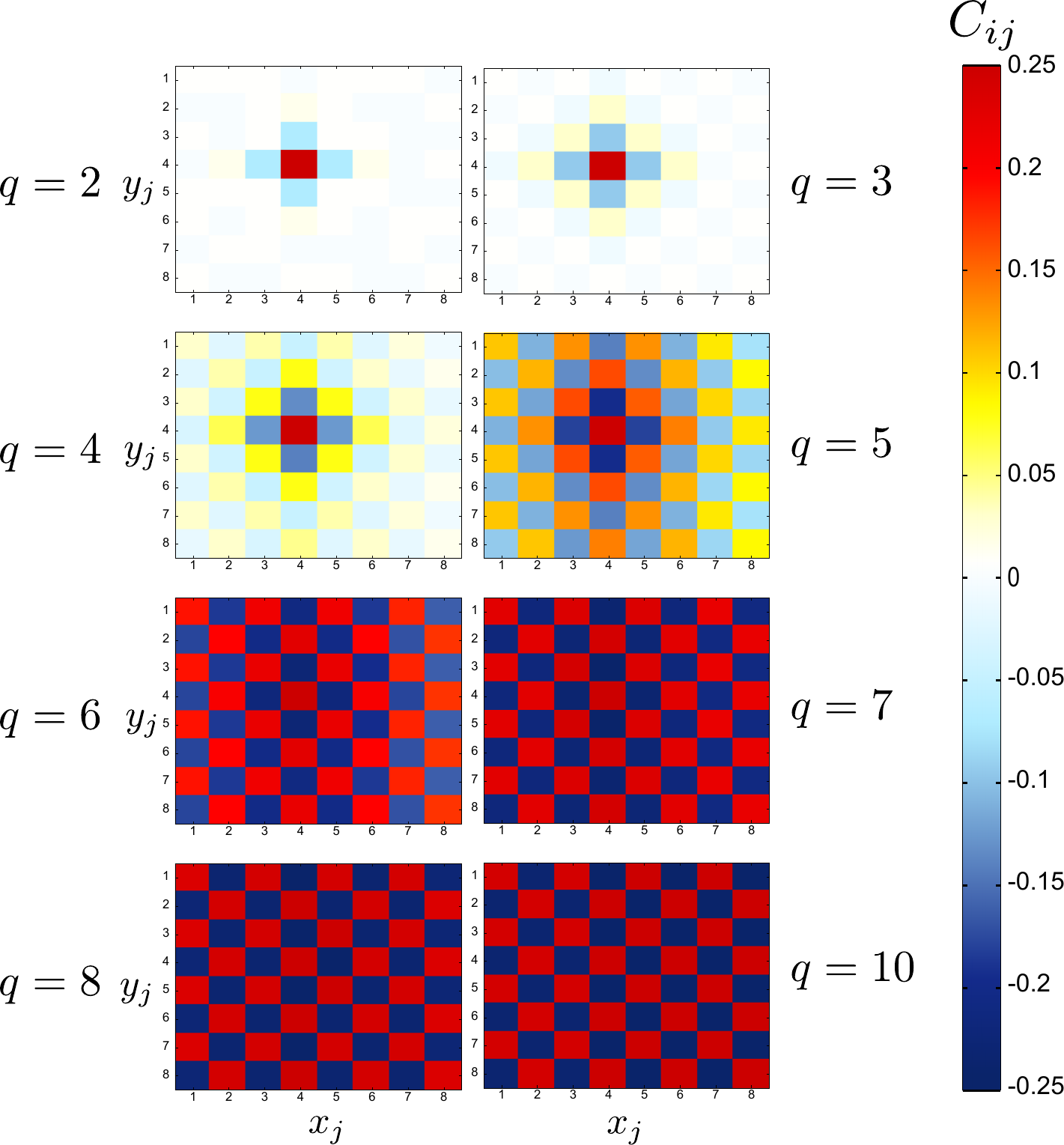}}

\subfigure[]{
        \centering
        \includegraphics[scale=0.65]{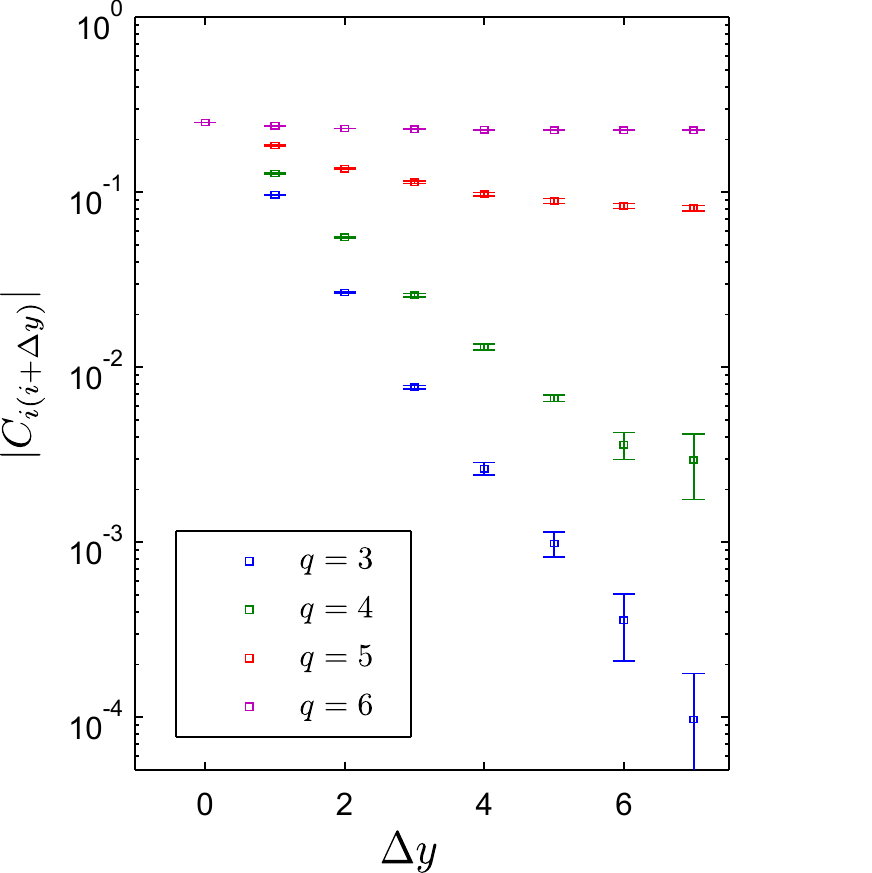}}

\caption{\label{corr2Dsquare}
(a) Correlation function $C_{ij}$ between a fixed lattice site $i$ in the middle of the lattice and all other lattice sites $j$ in the state $\psi_q^{q/2}$ on a square $8\times 8$ lattice on the cylinder.\\
(b) Absolute value of the correlation function $|C_{i(i+\Delta y)}|$ between two lattice sites separated by a distance $\Delta y$ in the periodic direction in the state $\psi_q^{q/2}$ on a square $16\times 16$ lattice on the cylinder. There is long range antiferromagnetic order for $q\geq 6$, while the correlations decay exponentially for $q\leq 4$.
}
\end{figure}

We first observe that for a given $q$ the correlation length decreases when $\eta$ goes away from the half-filling point $\eta=q/2$. At $\eta=1$, the correlation length also decreases when $q$ increases. At half-filling however, we find that the states $\psi_5^{2.5}$ and $\psi_6^{3}$ do not display exponentially decaying correlations, although for $\eta$ slightly different than $q/2$ the corresponding $\psi_q^\eta$ have exponentially decaying correlations. In Fig.~\ref{corr2Dsquare}, we show the decay of correlations in two dimensions (Fig.~\ref{corr2Dsquare}a) and along the periodic direction of the cylinder (Fig.~\ref{corr2Dsquare}b) for the half-filled states at different values of $q$. It is observed that for $q\leq 4$ the correlations decay exponentially. For $q\geq 6$, the states display clear long range anti-ferromagnetic correlations. The properties of the states at these particular points will be investigated in more details in Section \ref{section3}.  For $q=5$, the correlations decay and results on larger sizes suggest that long range anti-ferromagnetic order is also present at larger scales. In addition, we observe that Monte-Carlo simulations at $\eta=q/2$ need more computational effort to converge, in part due to the larger number of particles and in part due to the structure of the wave functions (see Section \ref{section3} for more details). The computations can be improved by exploiting the particle-hole symmetry at these points to allow global change of configurations respecting this symmetry in the Monte-Carlo paths.

\subsection{Topological entanglement entropy of the $\psi_q^\eta$ states}

\begin{figure}[htb]
\subfigure[]{
        \centering
        \includegraphics[scale=0.5]{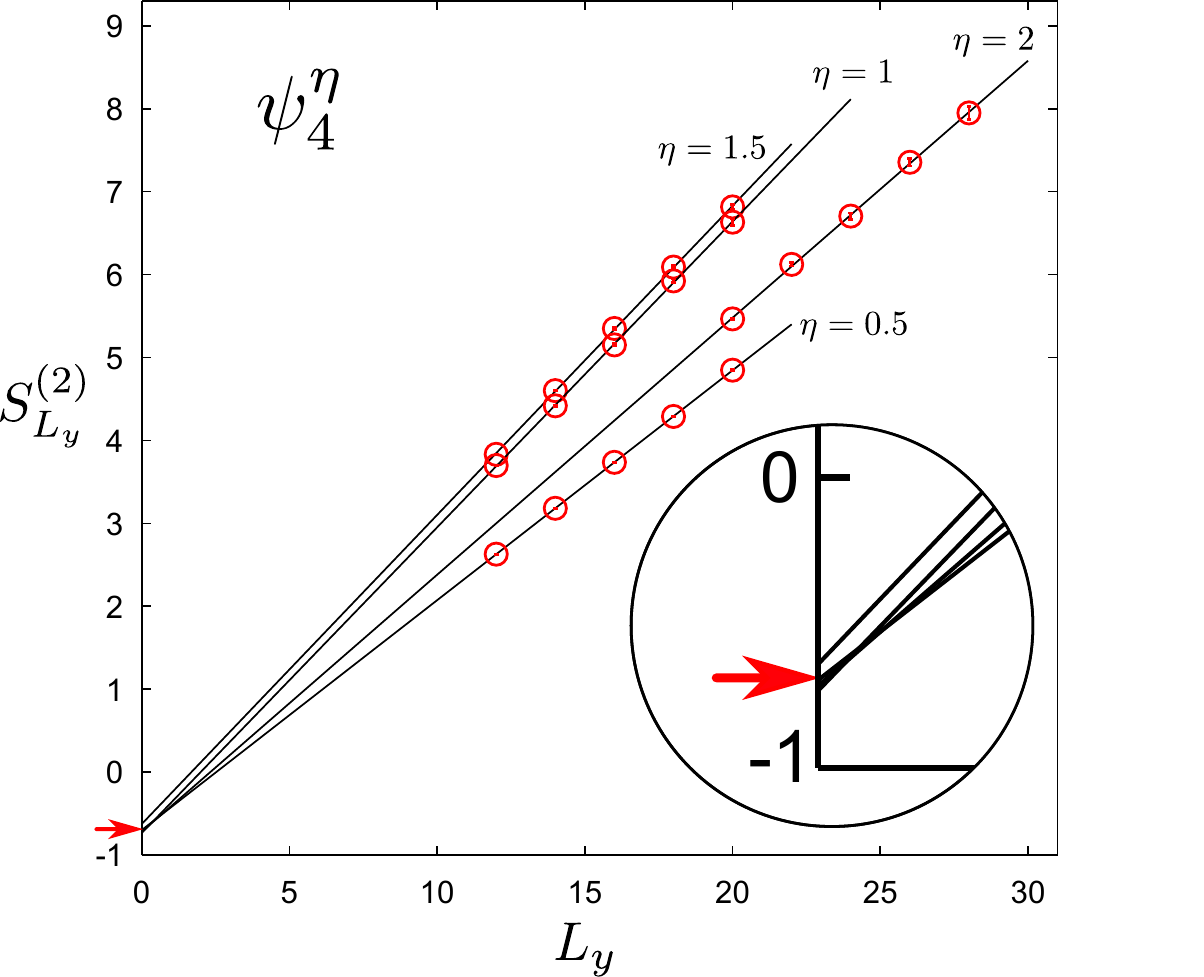}}

\subfigure[]{
        \centering
        \includegraphics[scale=0.5]{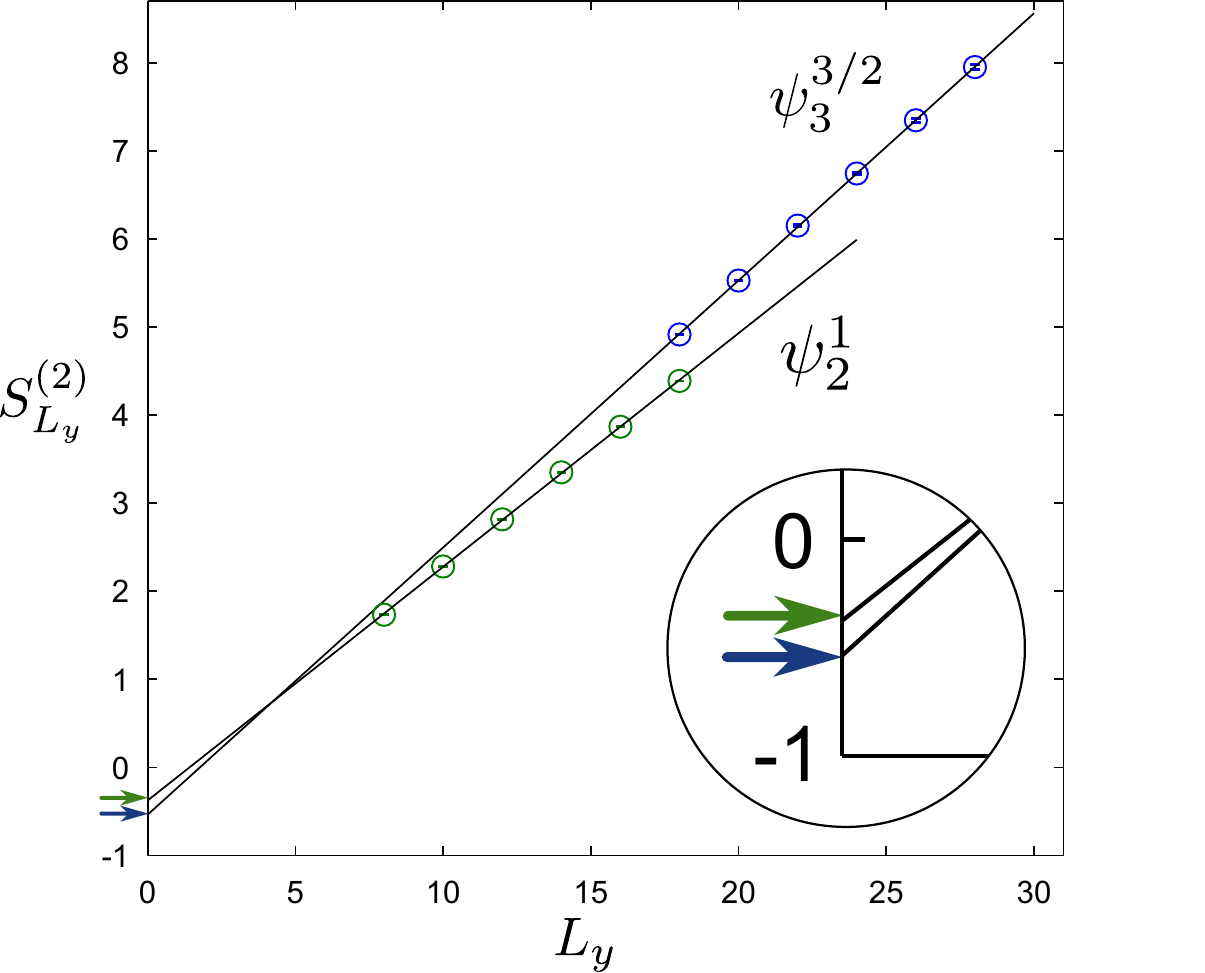}}
\caption{\label{teeq4}
(a) Scaling of the Renyi entropy $S^{(2)}_{L_y}$ of the state $\psi_4^\eta$ on a $L_x\times L_y$ square lattice on the cylinder (Fig.~\ref{fig:cylinder}). The topological entanglement entropy of the Laughlin state at filling $1/4$ ($\gamma_{0}(4)=\ln(2)\approx 0.693$) is indicated with a red arrow. The size $L_x$ is taken to be $12$, unless $\eta=2$ in which case $L_x$ is $20$. Larger sizes are taken when $\eta=q/2$, to account for the longer correlation length and to get rid of finite size effects. The black lines are linear fits and the values found for the topological entanglement entropy when $\eta$ equals to $0.5$, $1$, $1.5$ and $2$ are respectively $0.698$, $0.734$, $0.643$ and $0.718$.\\
(b) Scaling of the Renyi entropy $S^{(2)}_{L_y}$ for the states at half-filling $\psi_2^{1}$ and $\psi_3^{3/2}$ on the square lattice on the cylinder. Here $L_x$ is $12$ for $q=2$ and $20$ for $q=3$. The topological entanglement entropy of the Laughlin state at filling $1/2$, $\gamma_{0}(2)\approx 0.346$ (resp. at filling $1/3$, $\gamma_{0}(3)\approx 0.549$) is indicated with a green (resp. blue) arrow. The values found for the topological entanglement entropy are respectively $0.375$ and $0.536$.
}
\end{figure}

In the continuum, the Laughlin state at filling fraction $\nu=1/q$ has a topological entanglement entropy of \cite{Zozulya2007}
\begin{align}
\gamma_{0}(q)=\ln (\sqrt{q}).
\end{align}
To compute the topological entanglement entropy (TEE) of the CFT states, we start with a lattice on the cylinder, cut the cylinder into two halves and compute the Renyi entropy of the first half (Fig.~\ref{fig:cylinder}). The size along the cut is $L_y$ and we use the behaviour of the entanglement entropy \cite{Kitaev2006,Levin2006}
\begin{align}
\label{TEE}
S^{(2)}_{L_y}=\alpha L_y - \gamma
\end{align}
to extract the topological entanglement entropy $\gamma$, as has already been done using Monte-Carlo simulations for several chiral spin liquids for which the wave function is known\cite{Nielsen2012,Tu2014,Glasser2015,Wildeboer2015}. The results for the square lattice at $q=4$ and different values of $\eta$, shown in Fig.~\ref{teeq4}(a), confirm that the $\psi_4^\eta$ have the same topological entanglement entropy as the continuum Laughlin state at filling $1/4$, independently of the value of the lattice filling factor $\mu$. A particularly interesting case is the state $\psi_4^{2}$, defined on a half-filled lattice : this state is particle-hole symmetric, but has the topological entanglement entropy of the continuum Laughlin state at filling fraction $\nu=1/4$, which itself is not particle-hole symmetric. The same observation remains true at $q=2$ and $q=3$, where the particle-hole symmetric lattice state have the same topological entanglement entropy as the continuum Laughlin state at $\nu=1/q$ (see Fig.~\ref{teeq4}(b)). However, as we have seen previously the states $\psi_q^{q/2}$ for $q$ larger than $5$ on the square lattice have long range anti-ferromagnetic correlations. They do not display non-trivial topological behaviour, as will be clear from more investigations in Section \ref{section3}.

\subsection{Statistics of quasiholes in the $\psi_q^\eta$ states}

The previous results show that the $\psi_q^\eta$ states have similar properties as the continuum Laughlin states. A further tool that can be used to characterize the topological nature of these states is the braiding properties of its anyonic excitations, which can be computed easily when the wave function as well as the quasihole excitations are known analytically. In order to do this, we construct wave functions for the $\psi_q^\eta$ states in the presence of localized quasihole excitations\cite{Nielsen2015,Nielsen2015b}. We numerically test the screening of quasiholes, which allows us to compute their braiding properties and to confirm that these quasiholes have the same statistics as the anyons in the continuum Laughlin states.

Let us first start by defining wave functions for quasiholes, following Ref.~\onlinecite{Nielsen2015}. Consider $Q$ quasiholes at positions $w_j$, $j\in\{1, 2, \ldots , Q\}$, with charge $p_j/q$, where $p_j$ is a positive integer. We denote $P=\sum_{i}^{Q}p_i/q$ the total charge of the quasiholes. At the position of each quasihole, we attach a vertex operator
\begin{align}
W_{p_j}(w_j)=:e^{i\frac{p_j}{\sqrt{q}}\phi (w_{j})}:
\end{align}
and we consider the wavefunction given by 
\begin{align}
\label{correlator2}
\psi_q^\eta[& w_1,\ldots,w_Q](n_1, \ldots, n_N)\propto \nonumber\\
&\Braket{W_{p_1}(w_1)\dots W_{p_Q}(w_Q) V_{n_1}(z_1) \dots V_{n_N}(z_N)}.
\end{align}
This expression evaluates to \cite{Nielsen2015}
\begin{align}
\label{quasiparticlestate}
&\psi_q^\eta[p_1(w_1),\ldots,p_Q(w_Q)](n_1, \ldots, n_N)\propto \nonumber\\
&\delta'_n \xi_{\eta} \prod_{i,j}(w_i-z_j)^{p_i n_j}\prod_{i<j}^{N} (z_i-z_j)^{qn_i n_j}\prod_{i\neq j}^{N} (z_i-z_j)^{-\eta n_i} \nonumber\\
&\times \prod_{i<j}^{N} (w_i-w_j)^{p_i p_j /q} \prod_{i,j}(w_i-z_j)^{-p_i/q},
\end{align}
where $\delta'_n$ is zero unless the number of particles is 
\begin{align}
M=\sum_i n_i = \eta \frac{N}{q} - P.
\end{align}
In the thermodynamic limit, it can be shown\cite{Nielsen2015,Nielsen2015b} that this wave function for quasiholes is an analog of the wave function for quasiholes of Laughlin states in the continuum\cite{Moore1991}, but with the position of the particles restricted to the lattice sites. 

It was observed\cite{Nielsen2015,Nielsen2015b} that the anyonic statistics of the quasiholes can be computed by looking at the density of the state. Indeed, when the change of density due to the quasiholes is localized around the position of the quasiholes, then the Berry phase is zero\cite{Nielsen2015,Nielsen2015b} and therefore the statistics is governed by the monodromy, i.e. the change obtained from analytical continuation of the wave function when the quasiholes move around each other. For two quasiholes of charge $p_i$ and $p_j$ when the $i_{th}$ quasihole is moved in the counter-clockwise direction around the $j_{th}$ quasihole, the monodromy is $e^{2\pi i p_i p_j/q}$, which corresponds to the statistics expected for quasihole excitations of the Laughlin state at $\nu=1/q$.

\begin{figure}[htb]
\includegraphics[scale=0.64]{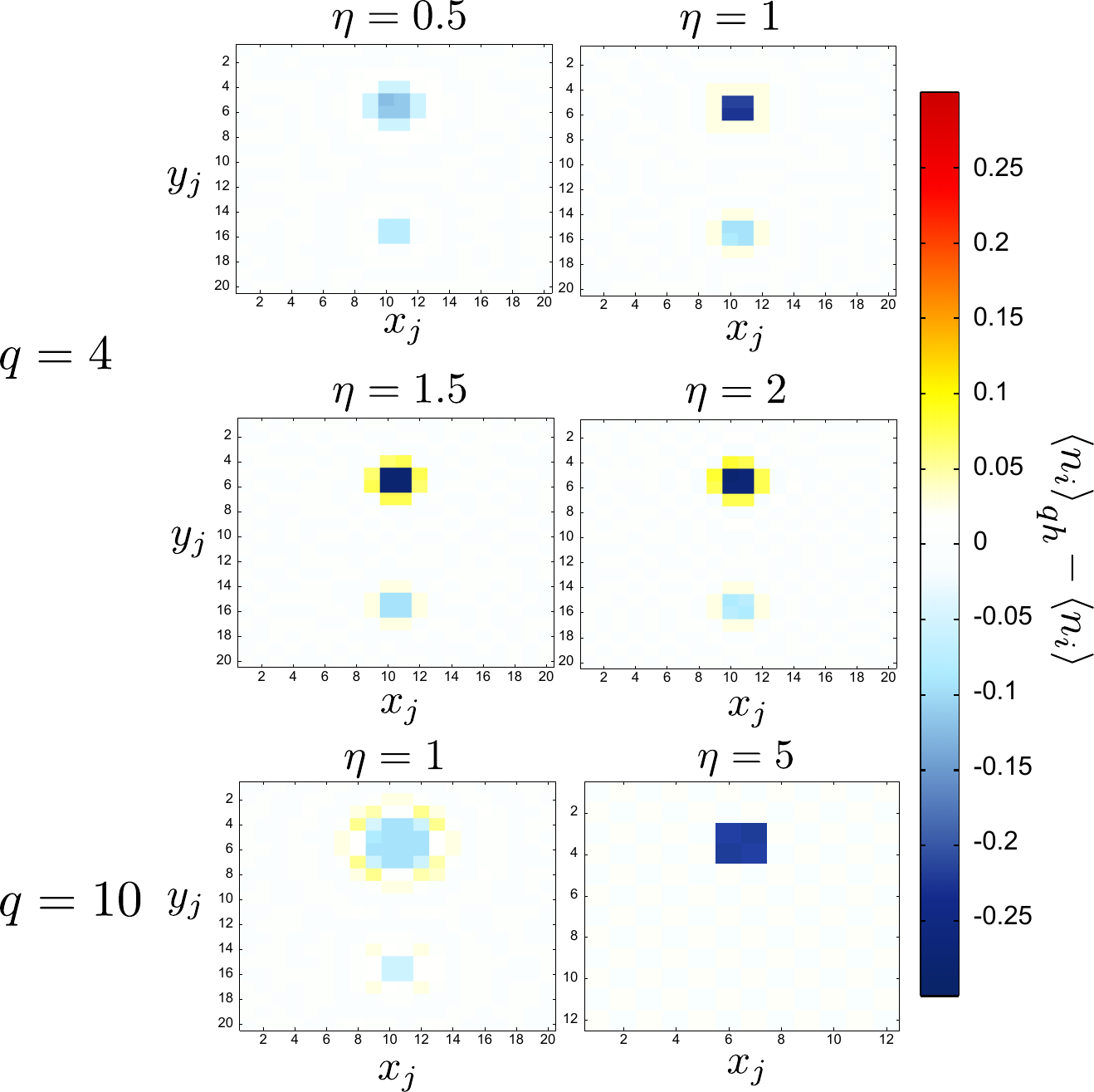}
\caption{\label{locquasi}
  Difference in the density $\Braket{n_i}_{qh}-\Braket{n_i}$ between the state $\psi_q^\eta$ with one quasihole of charge $1/q$ and one quasihole of charge $(q-1)/q$ and the state $\psi_q^\eta$ without quasiholes on a square $20\times 20$ (or $12\times 12$ for $q=10$, $\eta=5$) lattice on the cylinder. The coordinates $w_j$ of the quasiholes are placed in the center between $4$ lattice sites ($x_j=10.5$, $y_j=5.5$ for the quasihole with charge $(q-1)/q$ and $x_j=10.5$, $y_j=15.5$ for the quasihole with charge $1/q$) and are visible in blue as a lack of density on the neighbouring sites. At $q=4$, it is found for all values of $\eta$ that the quasiholes are localized. For $q=10$ however, the quasiholes are localized when $\eta=1$, but at half-filling ($\eta=5$) we observe that there is no splitting of the charge between a quasihole of charge $1/q$ and a quasihole of charge $(q-1)/q$ and thus no screening of the quasiholes.
}
\end{figure}

It therefore remains to be checked whether the change of density due to the presence of a quasihole is localized for any value of $\eta$. In Fig.~\ref{locquasi} we give numerical evidence that this is the case for different values of $\eta$ at $q=4$. We also show that the quasiholes are localized even for larger values of $q$ (at least up to $q=10$) at $\eta=1$, so that these wave functions may describe lattice Laughlin states for such $q$. However at $q=10$ and half-filling of the lattice ($\eta=5$), we find that there is no screening of the quasihole of charge $1/q$. This is an additional indication that for large $q$ and at half-filling of the lattice the states do not display the same topological properties as for lower fillings.

\section{States at half-filling : from long-range order on the square lattice to topological order on frustrated lattices}
\label{section3}

So far we have investigated the properties of the states on a square lattice. As we have seen, the states on this lattice are lattice versions of the continuum Laughlin states, except at half-filling when $q\geq 5$ where the states display long-range anti-ferromagnetic order. We now focus on these particular values of the parameters and investigate this effect in more details. We show that on the square lattice these states are well described by a simple combination of two N\'{e}el states that explain the behaviour of the correlations. We then turn to frustrated lattices and show that long-range order is not present on these lattices and that the expected topological behaviour is recovered.

\subsection{Long range antiferromagnetic order and N\'{e}el states}

The correlations previously computed show that for large values of $q$ and $\eta=q/2$, the state on the square lattice has long range anti-ferromagnetic correlations  reminiscent of N\'{e}el states. To understand this, let us look at the behaviour in the limit of infinite $q$. The dominating term in the wave function has the form $\prod_{i<j}(z_i-z_j)^{qn_i n_j}$, and only terms where the two positions $z_i$ and $z_j$ are occupied by a particle contribute to the wave function. The main contribution to the wave function therefore comes from states that maximize the product of the distances between pairs of particles on the lattice. In the infinite $q$ limit, only the states maximizing this product contribute to the wave function and the contribution of the other terms is suppressed. Since the lattice is half-filled, it is not possible to put particles too far apart to maximize this product. For the square lattice, the maximum is obtained when lattice sites alternate between empty and occupied sites in a checkerboard pattern. There are two such possibilities, corresponding to the two N\'{e}el states, that we denote $\psi_\text{N\'{e}el}^1$ and $\psi_\text{N\'{e}el}^2$ and that are related by particle-hole transformation.

\begin{figure}[htb]
\includegraphics[scale=0.55]{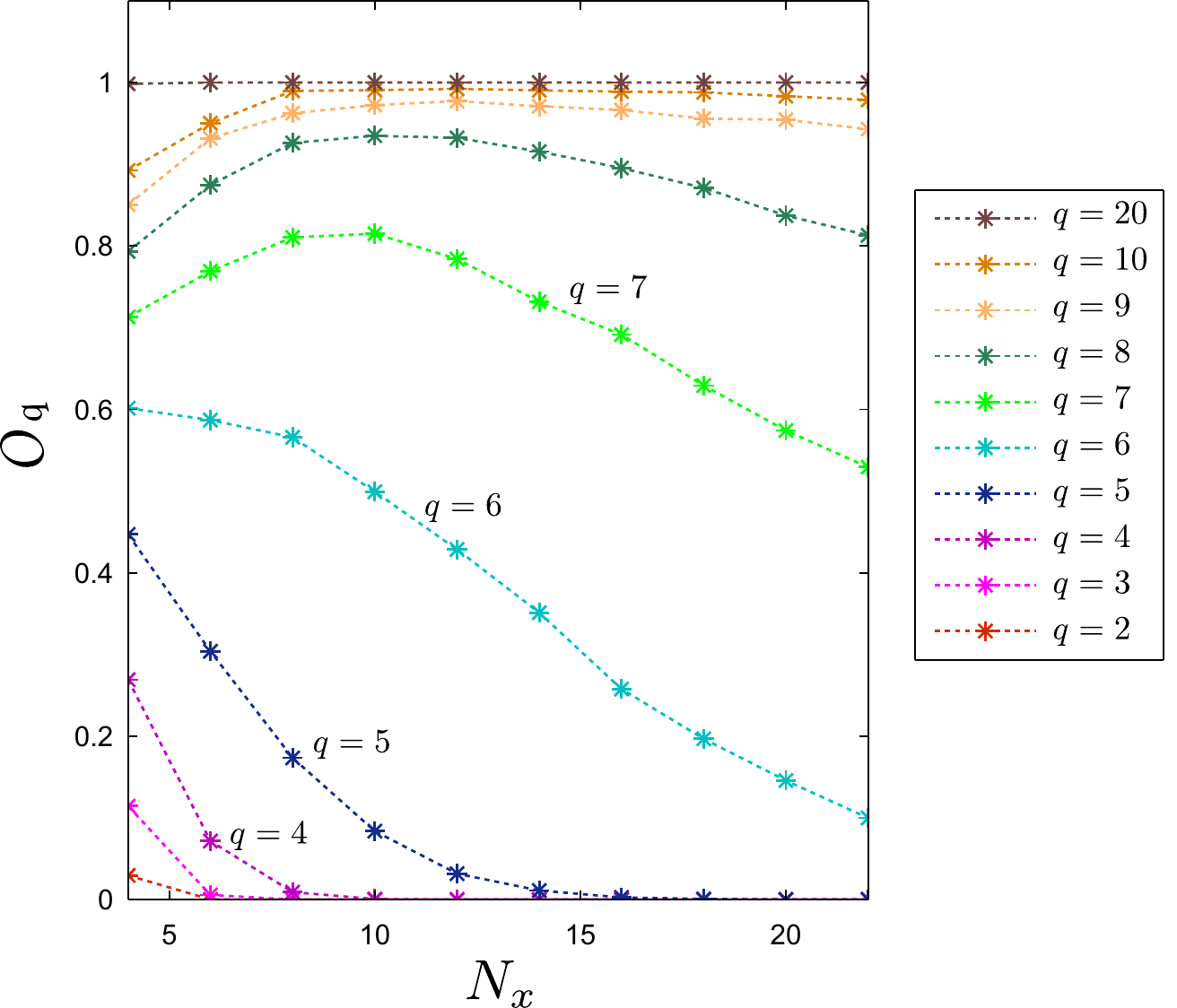}
\caption{\label{overlapantiferromagnet}
  Overlap, defined as $O_q\equiv |\Braket{\psi_\text{N\'{e}el}^1|\psi_q^{q/2}}|^2+|\Braket{\psi_\text{N\'{e}el}^2|\psi_q^{q/2}}|^2$, between the state at half-filling $\psi_q^{q/2}$ and the two N\'{e}el states, computed using a Metropolis-Hastings algorithm on a square lattice of size $N_x\times N_x$ on the cylinder. The errors from the Monte Carlo simulations are below $0.02$ for all points.
}
\end{figure}

To investigate how much these two N\'{e}el states contribute to the wave function, we compute the overlaps between each N\'{e}el state and the state $\psi_q^{q/2}$ for different lattice sizes (see Fig.~\ref{overlapantiferromagnet}). For large values of $q$, it is found that $O_q\equiv |\Braket{\psi_\text{N\'{e}el}^1|\psi_q^{q/2}}|^2+|\Braket{\psi_\text{N\'{e}el}^2|\psi_q^{q/2}}|^2$ is close to one, which shows that the wave function is almost a simple superposition of $\psi_\text{N\'{e}el}^1$ and $\psi_\text{N\'{e}el}^2$. For smaller values of $q$, this quantity goes rapidly to zero, while there is a transition in the range $5\leq q \leq 7$ where the overlap goes to zero but remains high, especially considering the size of the Hilbert space for the lattices considered. This explains the behaviour of the correlations, which decay exponentially for small $q$ but have long range order for large $q$. For other bipartite lattices which can support N\'{e}el states the argument given for infinite $q$ remains valid, so we expect a behaviour similar as for the square lattice. However the transition range might happen for different values of $q$.

\subsection{Frustration destroys the antiferromagnetic order}

\begin{figure}[htb]
\subfigure[]{
        \centering
        \includegraphics[scale=0.52]{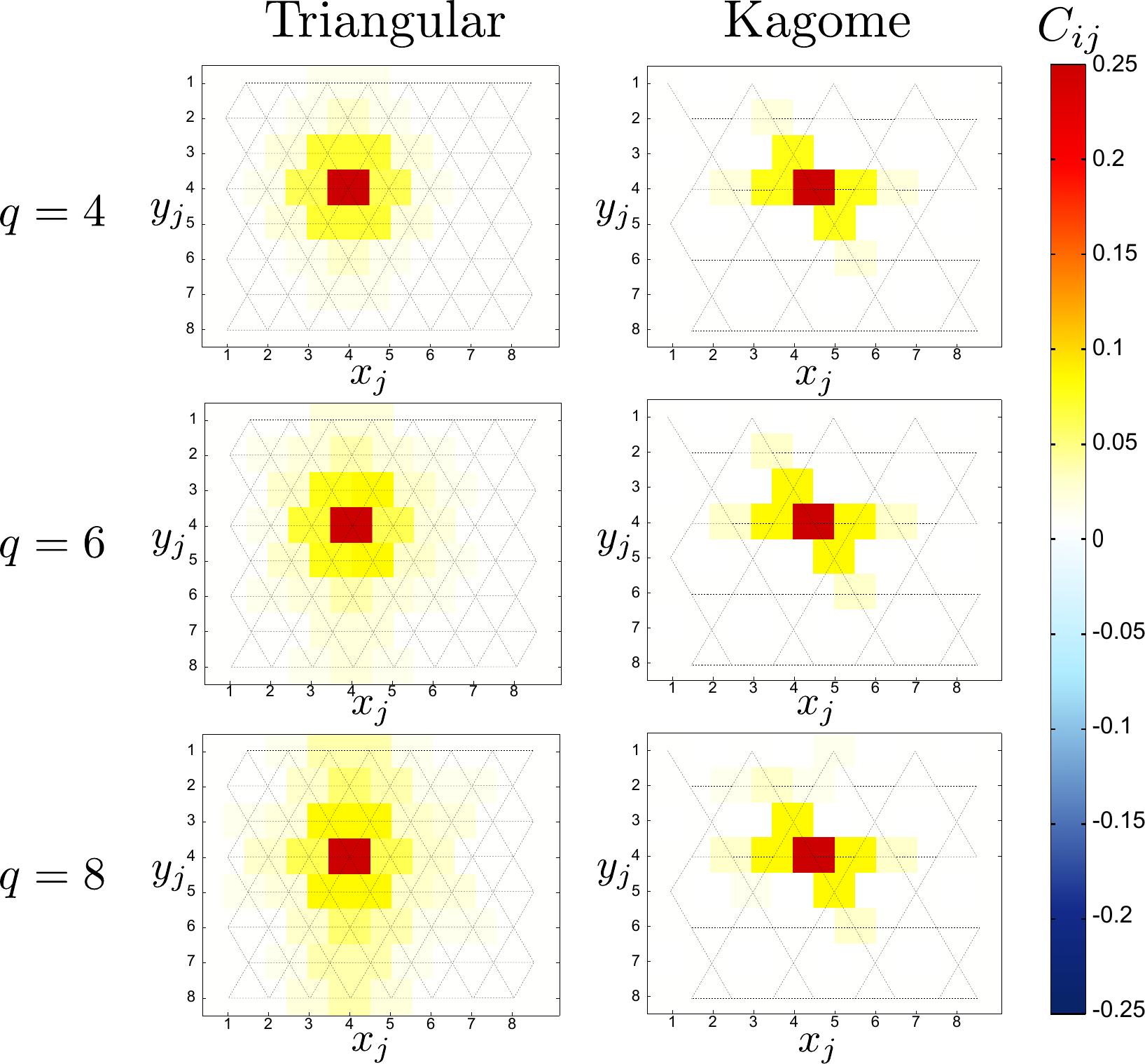}}

\subfigure[]{
        \centering
        \includegraphics[scale=0.65]{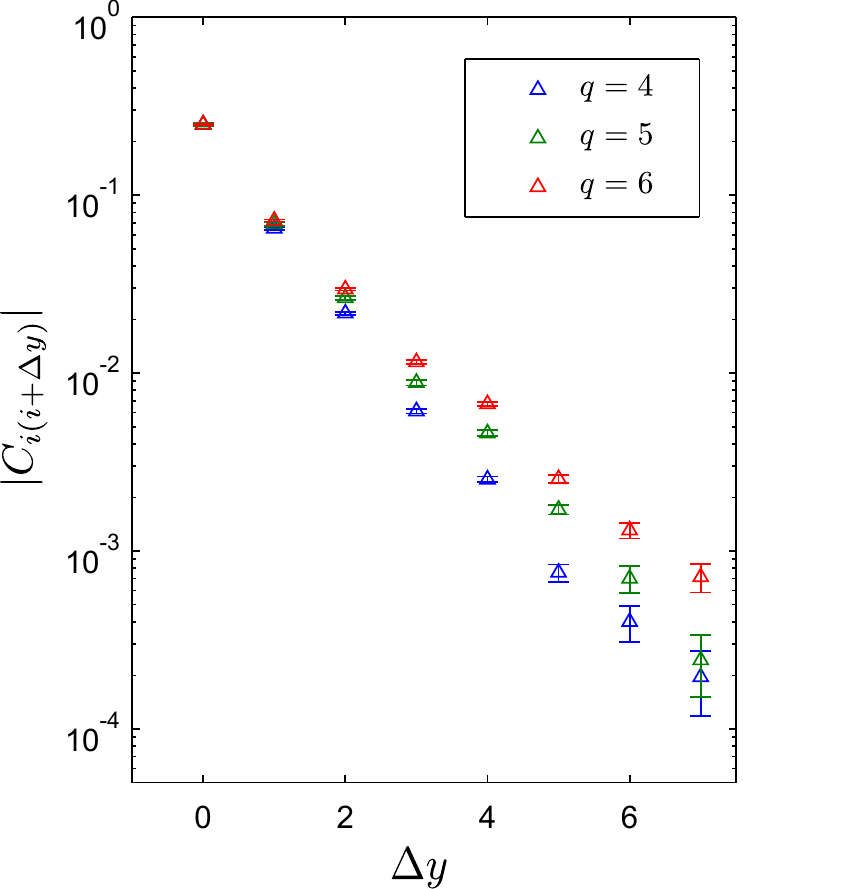}}
\caption{\label{corr2Dtriangle}
(a) Correlation function $C_{ij}$ between a fixed lattice site $i$ in the middle of the lattice and all other lattice sites $j$ in the state $\psi_q^{q/2}$ on a triangular and Kagome lattice on the cylinder.\\
(b) Absolute value of the correlation function $|C_{i(i+\Delta y)}|$ between two lattice sites separated by a distance $\Delta y$ in the periodic direction in the state $\psi_q^{q/2}$ on a triangular $16\times 16$ lattice on the cylinder.\\
Unlike on the square lattice for the same parameters, there is no long range order.
}
\end{figure}

Let us note that the two N\'{e}el states can arise because of the geometry of the square lattice. In this particular case, strong lattice effects at half-filling can give rise to antiferromagnetic behaviour. On frustrated lattices, we cannot define two N\'{e}el states respecting the symmetries of the lattice. We therefore now turn on to investigations of the $\psi_q^{q/2}$ on the triangular and Kagome lattices. Fig.~\ref{corr2Dtriangle} shows evidence that the particle-particle correlations on the triangular and Kagome lattices decay exponentially with the distance even for $q=6$, in sharp contrast with the behavior on the square lattice.

\begin{figure}[htb]
        \centering
        \includegraphics[scale=0.46]{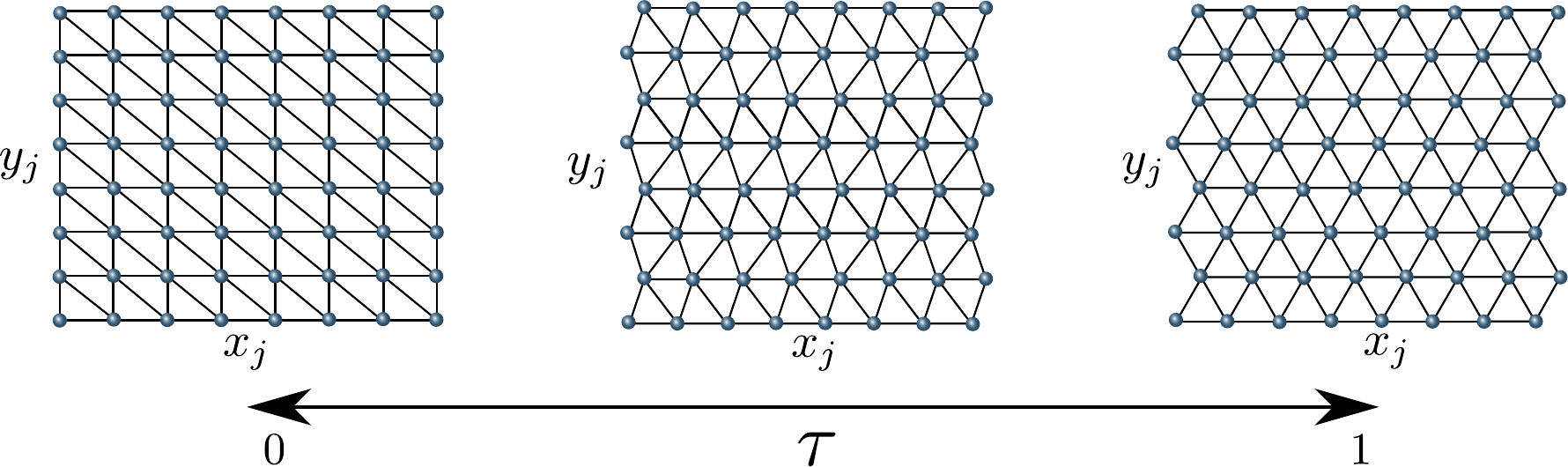}
\caption{\label{interpotriangle}
Positions of the coordinates of the lattice sites along the interpolation between a square and a triangular lattice. The coordinates along the periodic direction $y_j$ are kept fixed to keep the periodicity, while the coordinates along the other direction $x_j$ are linearly interpolated between the square ($\tau=0$) and the triangular ($\tau=1$) lattice. The coordinates on the plane are then $z_j=e^{\frac{2\pi}{L_y} (x_j+iy_j)}$, as in Fig.~\ref{fig:cylinder}.
}
\end{figure}

\begin{figure}[htb]
        \centering
        \includegraphics[scale=0.5]{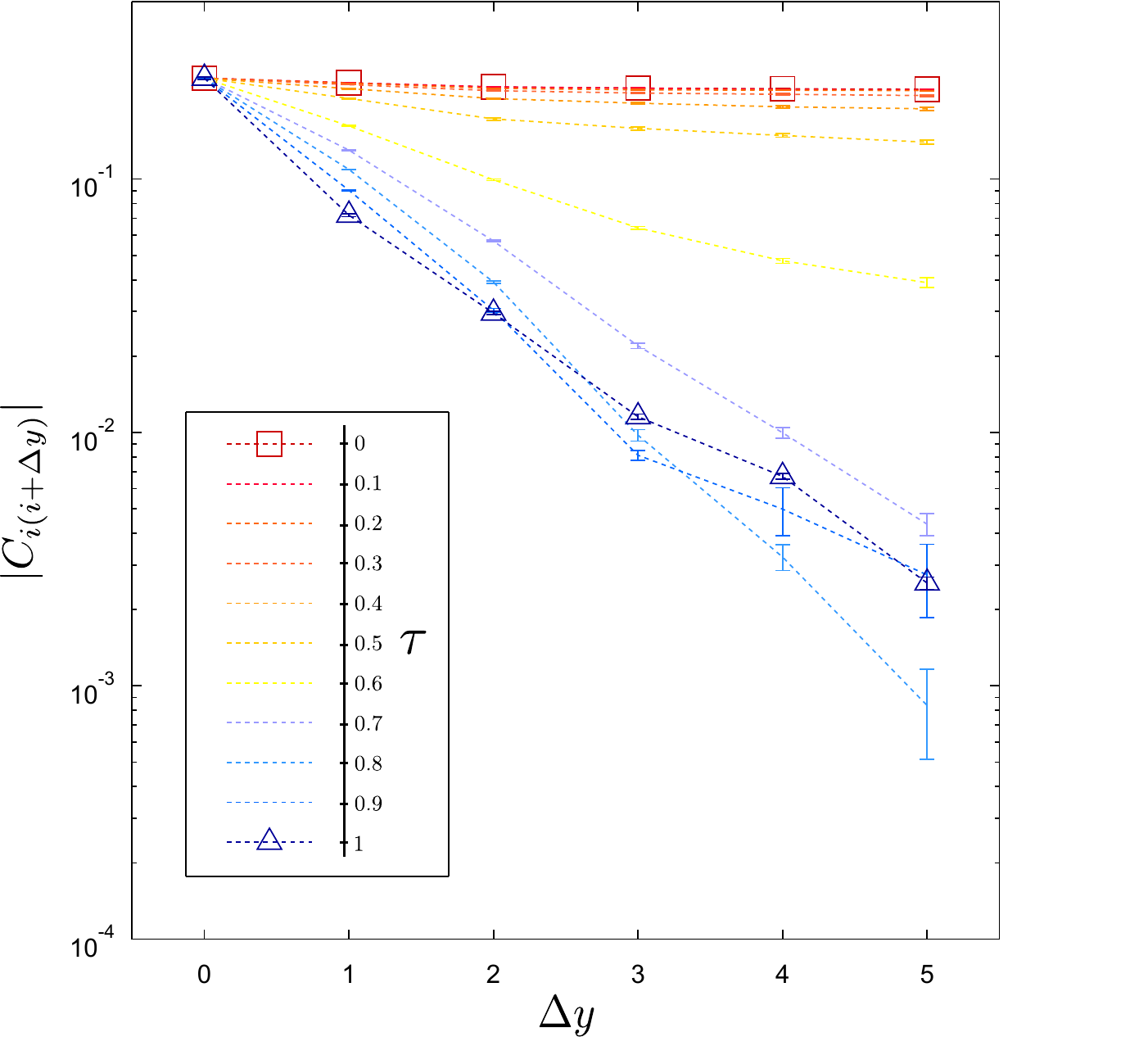}
\caption{\label{interpotrianglecorr}
Absolute value of the correlation function $|C_{i(i+\Delta y)}|$ between two lattice sites separated by a distance $\Delta y$ in the periodic direction in the state $\psi_6^{3}$ on a $16\times 16$ lattice interpolating between the square ($\tau=0$) and the triangular ($\tau=1$) lattice on the cylinder.
}
\end{figure}

Since the states are defined as a function of the positions $z_i$ of the lattice points, we can study the transition between the square and the triangular lattice by changing continuously the coordinates $z_i$ to interpolate linearly between a square lattice and a triangular lattice on the cylinder (Fig.~\ref{interpotriangle}). The decay of correlations for $q=6$ is shown in Fig.~\ref{interpotrianglecorr} for different values of the interpolation parameter $\tau$. We observe that the correlations decay exponentially when the lattice is close to a triangular lattice ($0.7\leq\tau\leq 1$), while there is a transition towards anti-ferromagnetic order when the lattice gets closer to the square lattice. This confirms the importance of the geometry of the lattice in the study of the wave functions $\psi_q^{q/2}$.

\subsection{Topological entanglement entropy on the triangular lattice}

\begin{figure}[htb]
        \centering
        \includegraphics[scale=0.5]{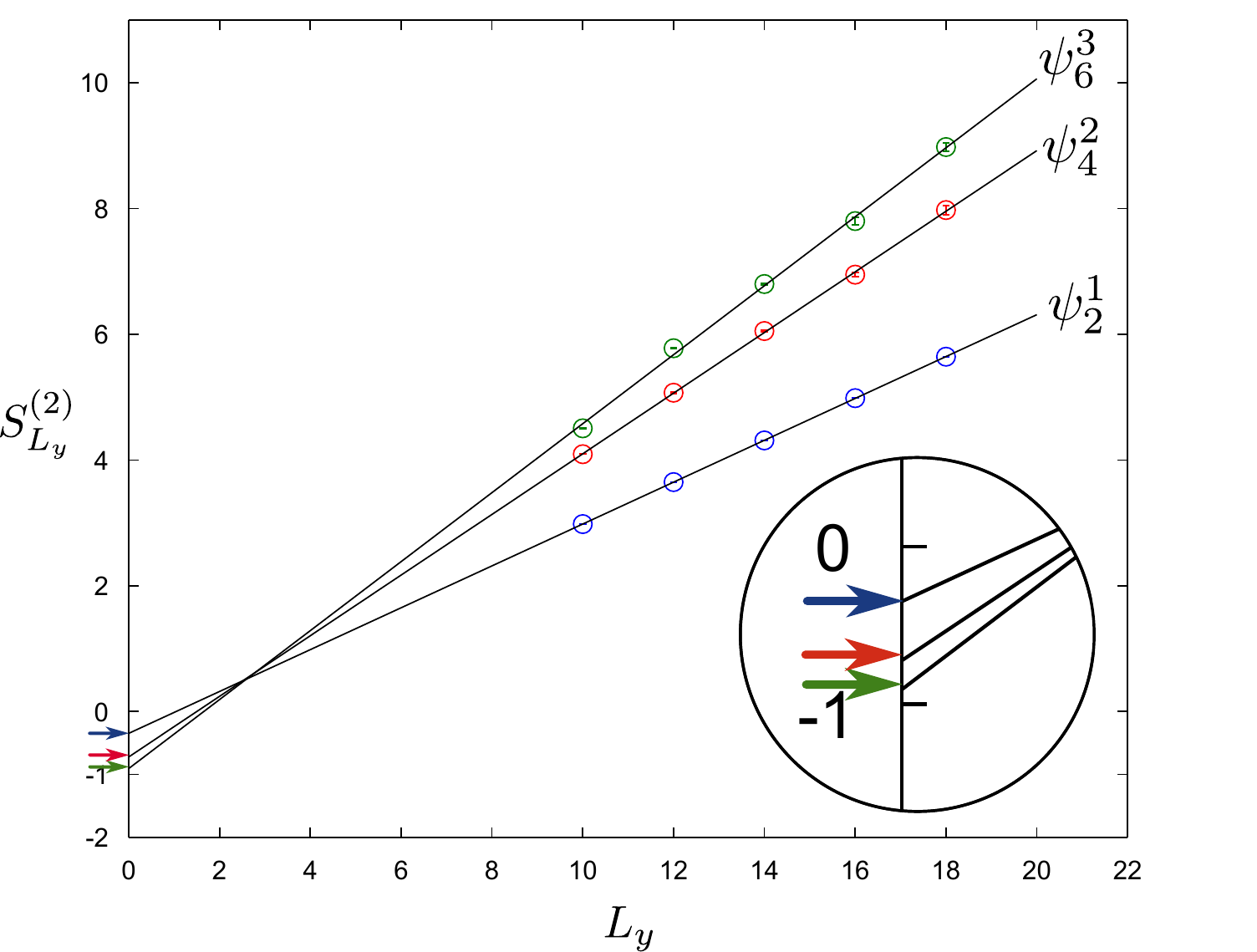}
\caption{\label{teeqtriangle}
(b) Scaling of the Renyi entropy $S^{(2)}_{L_y}$ for the states at half-filling $\psi_2^{1}$, $\psi_4^{2}$ and $\psi_6^{2}$ on a triangular lattice on the cylinder. Here $L_x$ is taken to be $16$. The topological entanglement entropy of the continuum Laughlin state at $q=2$ (resp. $q=4$, $q=6$) is indicated with a blue (resp. red, green) arrow. The values found for the topological entanglement entropy are respectively 0.347, 0.723 and 0.907.
}
\end{figure}

It was shown previously that the long range order was destroyed on frustrated lattices. We might therefore wonder whether these half-filled states on the triangular lattice have the same topological properties as the continuum Laughlin states at filling fraction $\nu=1/q$. We compute in Fig.~\ref{teeqtriangle} the topological entanglement entropy of the state $\psi_q^{q/2}$ on the triangular lattice and show that its value is compatible with the value for the corresponding continuum Laughlin states at $q=2, 4$ and $6$. Particle-hole symmetric lattice Laughlin states with topological order on the triangular lattice may therefore be described by these wave functions.  

\section{Edge states from a charge at infinity}
\label{sectionedge}

In this section we define wave functions for edge states of the lattice Laughlin states. These states are constructed using a charge operator placed at infinity and their wave functions have the same expression as the lattice Laughlin states, except that the number of particles is different. They have the same correlations as the $\psi_q^\eta$ states in the bulk but a different density at the edge, as we shall see below.\\

\begin{figure}[htb]
\subfigure[]{
        \centering
        \includegraphics[scale=0.52]{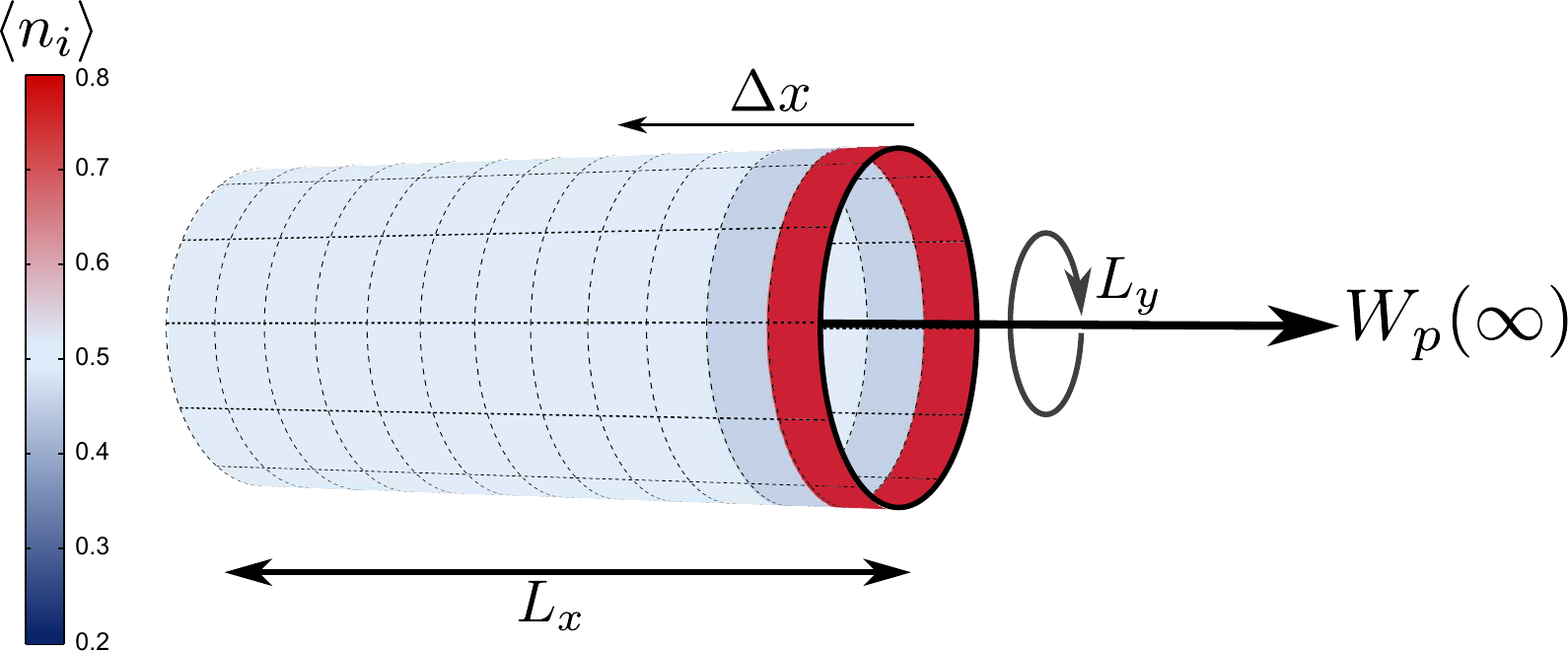}}

\subfigure[]{
        \centering
        \includegraphics[scale=0.52]{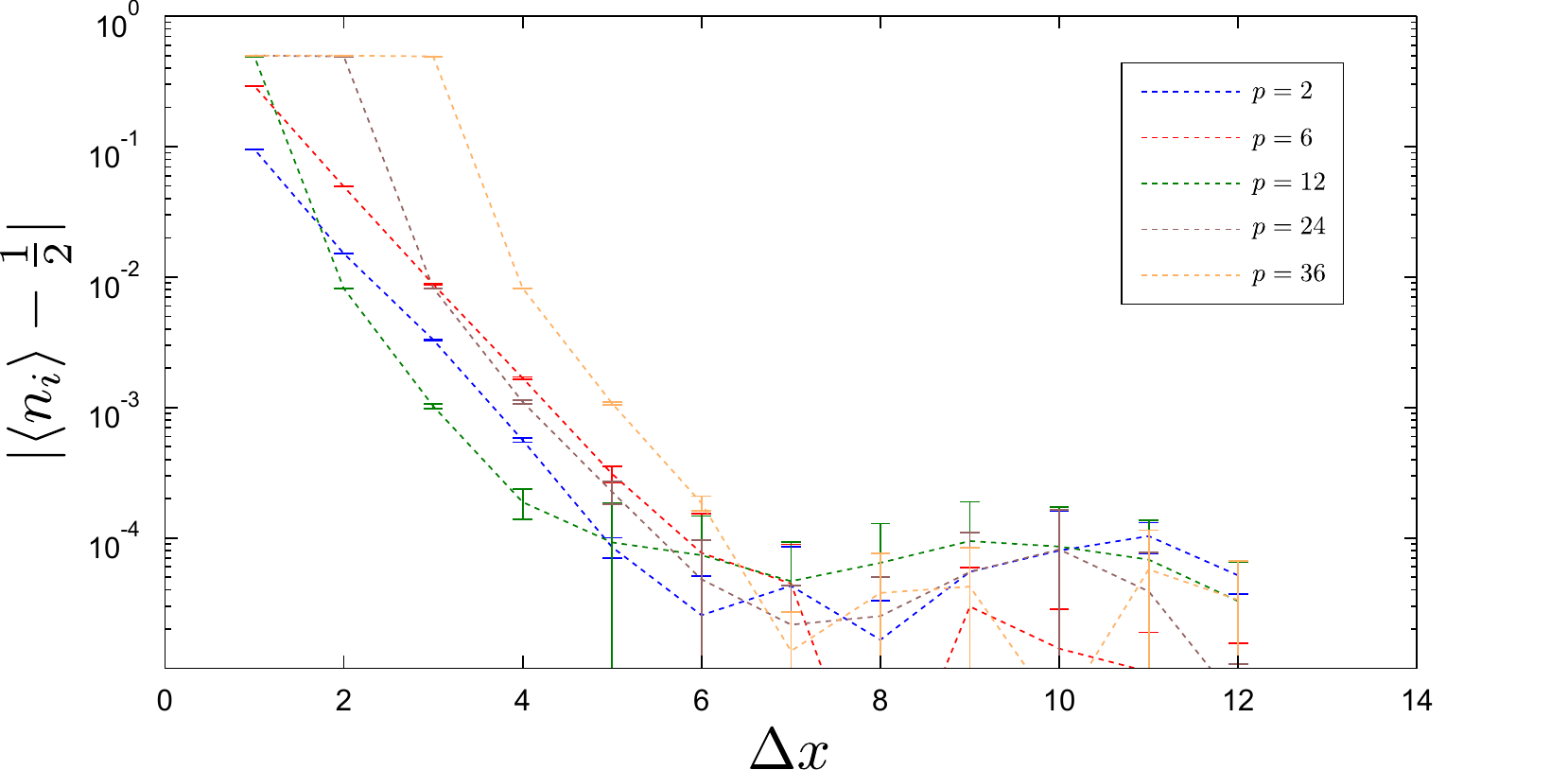}}

\caption{\label{fig:edgedensity}
(a) Square $12 \times 12$ lattice on the cylinder with a charge (here $p=-6$) at infinity. The colors show the density of the state $\psi_2^1[p]_\infty$. The state with zero charge at infinity $\psi_2^1$ has uniform density $1/2$ but here the density is modified at the edge to account for the change of total particle number.\\
(b) Difference of density between the $\psi_2^1[p]_\infty$ states and the $\psi_2^1$ state with respect to the distance to the edge $\Delta x$. Values below $2\times 10^{-4}$ are not converged. The change of density is exponentially localized at the edge.
}
\end{figure}

As we have seen previously, the presence of quasiholes affects the number of particles in the lattice, and hence also the filling factor $\mu$ of the lattice. However, since the quasiholes are localized, the density is affected locally and far from any quasihole the density stays equal to $\mu$. Let us investigate the effect of inserting in the correlator defining the wave function a single operator $W_{p}(\infty)=:e^{i\frac{p}{\sqrt{q}}\phi (\infty)}:$ of charge $p/q$, $p$ integer, when the position of the operator is taken to infinity. When the lattice is mapped to the cylinder, this amounts to having a charge at infinity on the axis of the cylinder (Fig.~\ref{fig:edgedensity}). We suppose here that we start with a state at $\eta=1$, so the wave function is  
\begin{align}
\label{quasiparticleedgestate}
\psi_q^1&[p(\infty)]\propto {\underset{w\rightarrow \infty}\lim} \delta'_n \prod_{j}(w-z_j)^{p n_j}\prod_{i<j}^{N} (z_i-z_j)^{qn_i n_j}\nonumber\\
&\qquad \qquad \times \prod_{i\neq j}^{N} (z_i-z_j)^{-n_i},\nonumber\\
&\propto {\underset{w\rightarrow \infty}\lim} \delta'_n w^{p(N-p)/q} \prod_{i<j}^{N} (z_i-z_j)^{qn_i n_j} \prod_{i\neq j}^{N} (z_i-z_j)^{- n_i},\nonumber\\
&\propto \delta'_n \prod_{i<j}^{N} (z_i-z_j)^{qn_i n_j} \prod_{i\neq j}^{N} (z_i-z_j)^{- n_i},
\end{align}
where $\delta'_n$ is zero unless the number of particles is 
\begin{align}
M=\sum_i n_i = \frac{N-p}{q}.
\end{align}
We denote this state $\psi_q^1[p]_\infty$. It has exactly the same expression for the wave function as the state $\psi_q^1$, except that the number of particles has been modified to $\frac{N-p}{q}$. In the particular case $p=N(1-\eta)$, then the wave functions $\psi_q^1[p]_\infty$ and $\psi_q^\eta$ have the same lattice filling factor $\mu=\eta/q$ and differ only by an $\eta$ exponent in the last term of the wave function.

We show in Fig.~\ref{fig:edgedensity} that these two states differ in their density : the density of the $\psi_2^1[p]_\infty$ state on the cylinder is the same as the density of the $\psi_2^1$ state ($1/2$) in the bulk, but is modified on the edge to account for the different total number of particles. This modification of the density is exponentially decaying from the edge towards the bulk of the system. This is in contrast with the density of the $\psi_2^\eta$ state with the same number of particles, which has a roughly uniform density of $\eta/q$.

\begin{figure}[htb]
\includegraphics[scale=0.52]{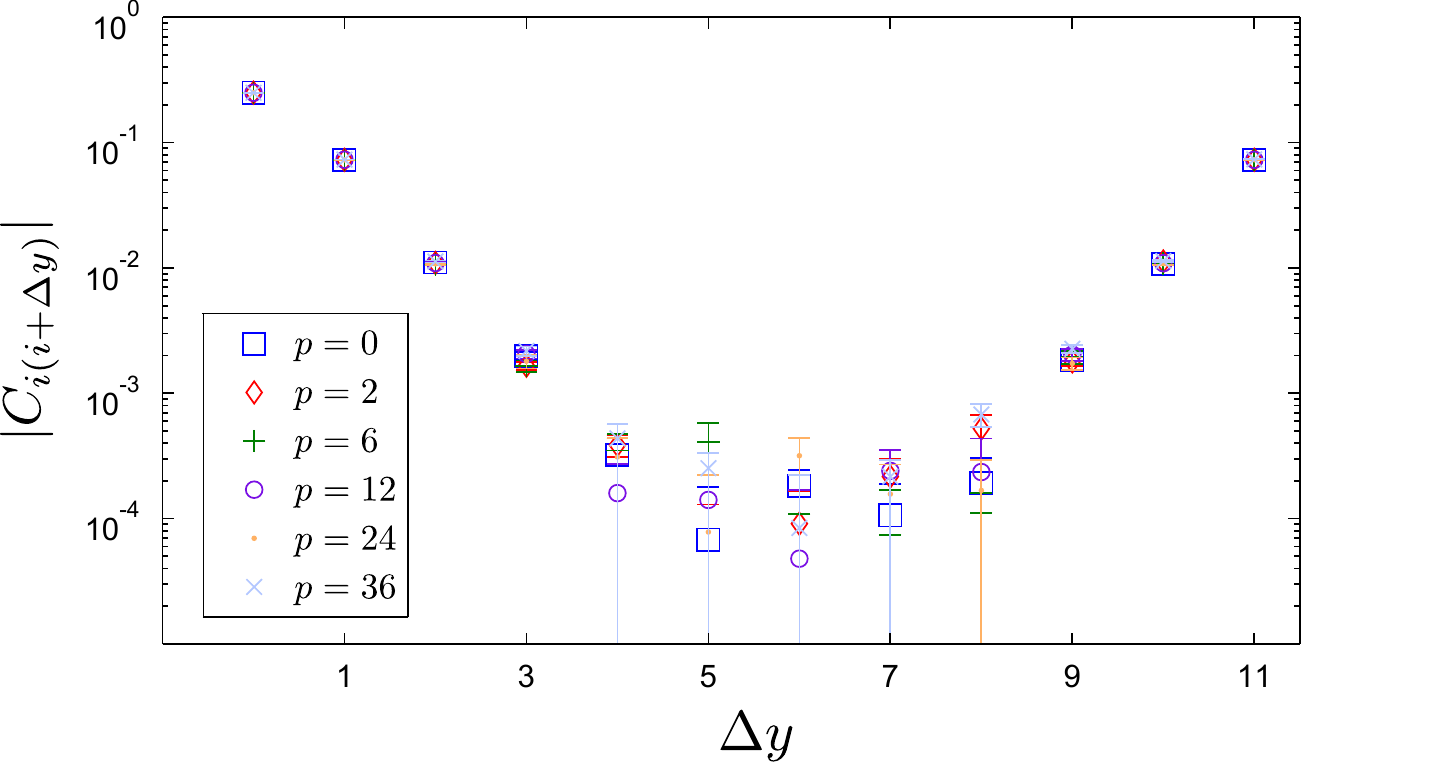}
\caption{\label{fig:edgecorrelations}
  Absolute value of the correlation function $|C_{i(i+\Delta y)}|$ between two lattice sites in the bulk (middle of the cylinder) separated by a distance $\Delta y$ in the periodic direction on a square $12\times 12$ lattice on the cylinder. The states considered here are the $\psi_q^{1}[p]_\infty$ state for different values of $p$. When $p=0$, the state is simply the $\psi_2^{1}$ state, while it is found that the other states have the same correlations.
}
\end{figure}

In addition, we give numerical evidence in Fig.~\ref{fig:edgecorrelations} that the states $\psi_2^1[p]_\infty$ and $\psi_2^1$ have the same particle-particle correlations in the bulk. Therefore the $\psi_2^1[p]_\infty$ states can be understood as edge states with respect to the $\psi_2^1$ wave function, since they differ from it only at the edge.

The properties investigated here are similar to the properties of the edge states defined in Ref.~\onlinecite{Herwerth2015}, but their construction and wave functions are different.

Note that on the cylinder it is also possible to describe edge states on the other edge by choosing the position of the charge to be $0$ instead of infinity. Since the coordinates of the lattice on the cylinder as well as the coefficients of the wave functions are invariant under a transformation $z\rightarrow 1/z$, this simply amounts to flipping the two sides of the cylinder and as such will not change the properties of the states.

\section{Derivation of parent Hamiltonians}

\label{sectionparentH}
\begin{figure}[htb]
\includegraphics[scale=0.6]{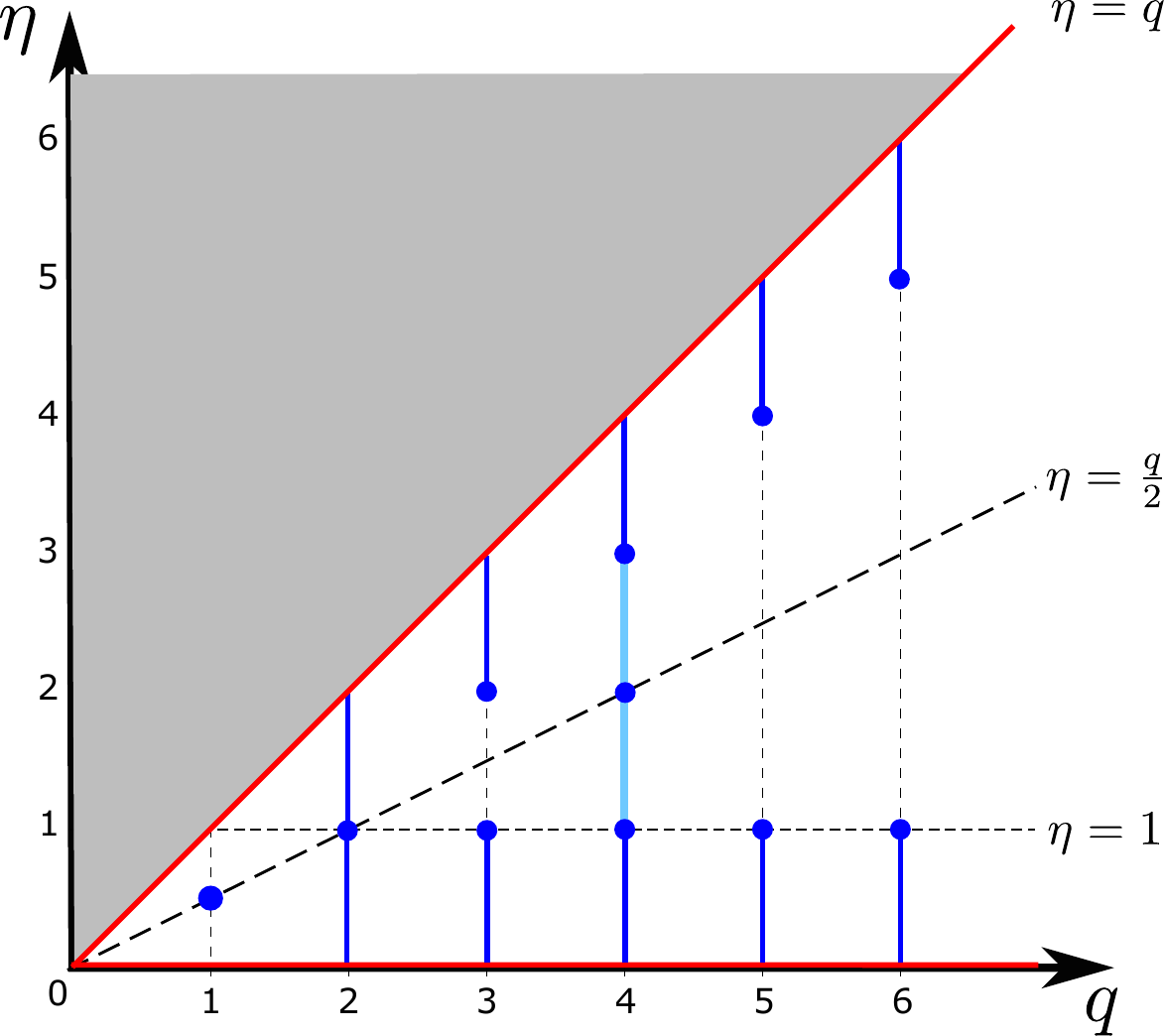}
\caption{\label{parenthamiltonian}
  Diagram of the $\psi_q^\eta$ states. The blue lines and dots represent values of the parameters for which exact parent Hamiltonians are derived in this section. The light blue line represents values of the parameters for which the parent Hamiltonians derived have a degenerate ground state on the plane.
}
\end{figure}

In this section we derive exact parent Hamiltonians of the lattice Laughlin states. We start by deriving a parent Hamiltonian for which the $\psi_q^1[p]_\infty$ edge states are ground states with different number of particles. The lattice Laughlin state $\psi_q^1$ is the unique ground state of this Hamiltonian with lattice filling factor $1/q$. We then provide a general method for deriving the parent Hamiltonian of a state when we already know the parent Hamiltonian for a different state which differs from the first only by a product of single-body operators. This allows us to obtain parent Hamiltonians of the $\psi_q^\eta$ states when $\eta\leq 1$ or $\eta\geq q-1$. Compared to the approach used to connect Hamiltonians in the lattice and in the continuum limit in Ref.~\onlinecite{Nielsen2015b}, our Hamiltonians do not increase in complexity when we go from the lattice to the continuum limit, since only the coupling strengths need to be changed with $\eta$. We finally consider the states at half-filling and derive a parent Hamiltonian for the $\psi_4^2$ state. Fig.\ref{parenthamiltonian} shows the diagram of the states and the values of the parameters for which parent Hamiltonians are obtained.

\subsection{Parent Hamiltonian for the edge state $\psi_q^1[p]_\infty$}
\label{parentH1}

Let us start by deriving parent Hamiltonians for the edge states $\psi_q^1[p]_\infty$.  This state is defined by introducing a charge operator in the wave function and taking the position of this operator going to infinity :
\begin{align}
\psi_q^1[p]_\infty(n_1, \ldots, n_N)\propto \Braket{W_{p}(\infty)V_{n_1}(z_1) \dots V_{n_N}(z_N)}.
\end{align}
In Ref.~\onlinecite{Nielsen2015}, an exact parent Hamiltonian was derived for states written in this form when $p>0$. We follow here the same procedure and extend this result to $p>-q$. The starting point is to consider a field $\chi(w)$ defined as
\begin{align}
\chi(w) &=\oint_{w}\frac{dz}{2\pi i}\frac{1}{z-w} [G^{+}(z)V_{-}(w) - qJ(z)V_{+}(w)],
\end{align}
where $G^{+}(z)=:e^{i\sqrt{q}\phi (z)}:$, $J(z)=i\partial_z \phi /\sqrt{q}$, $V_{-}(z)=V_{0}(z)$ and $V_{+}(z)=(-1)^{(j-1)}V_{1}(z)$ (here we assume $\eta=1$). $\chi(w)$ is a null field\cite{Tu2014,Nielsen2015}, so the correlator with this field inserted vanishes :
\begin{align}
\Braket{W_{p}(w)V_{n_1}(z_1) \dots V_{n_{i-1}}(z_{i-1}) \chi(z_i)\nonumber\\ \times V_{n_{i+1}}(z_{i+1}) \dots V_{n_N}(z_N)}=0.
\end{align}
As derived in more details in Appendix \ref{appendixA}, this equation can be rewritten, when $p>-q$, as
\begin{align}
\Lambda_i \Ket{\psi_q^1[p(w)]}=0,
\end{align}
where \begin{align}
\Lambda_i=\sum_{j(\neq i)} \frac{1}{z_i-z_j} [d_j-d_i (qn_j-1)] - \frac{p}{z_i-w} d_i.
\end{align}
Here $d_j$ is the hardcore boson (resp. fermion) annihilation operator for $q$ even (resp. odd), and $n_j=d_j^\dagger d_j$ is the number of particles at site $j$. Taking the limit $w\rightarrow \infty$, so that the charge is at infinity, leads to
\begin{align}
\label{eqlambdaedge}
\Lambda_i \Ket{\psi_q^1[p]_\infty}=0,
\end{align}
where 
\begin{align}
\Lambda_i=\sum_{j(\neq i)} \frac{1}{z_i-z_j} [d_j-d_i (qn_j-1)].
\end{align}
This leads to $H\Ket{\psi_q^1[p]_\infty}=0$, where 
\begin{align}
H=\sum_i \Lambda_i^\dagger \Lambda_i
\end{align}
is a positive semi-definite operator annihilating the wave function $\psi_q^1[p]_\infty$. Here we note that there is no dependence on $p$ in this Hamiltonian. This means that the Hamiltonian $H$ has a degenerate ground space containing all the $\psi_q^1[p]_\infty$ edge states for $p>-q$. These states have however different total number of particles, and in physical settings where the number of particles is fixed only one of these states would be ground state of the corresponding Hamiltonian. A particular ground state of this Hamiltonian is the state at $p=0$, which is the $\psi_q^1$ Laughlin state, corresponding to the ground state at filling factor $\mu=1/q$. Moreover we have checked numerically for small lattice sizes that $H$ has only one ground state for each subspace with fixed number of particles.

\subsection{Parent Hamiltonians for $\psi_q^\eta$ states}
\label{parentHeta}

Let us now turn to the derivation of parent Hamiltonians for the $\psi_q^\eta$ states. The first observation is that the $\psi_q^\eta$ states differ from the $\psi_q^1[p]_\infty$ states with $p=N(1-\eta)$ only by a factor $ \prod_{i\neq j}^{N} (z_i-z_j)^{(\eta-1)n_i}$. We will use our previous result and account for this factor at the level of the Hamiltonian. Let us denote $T$ the operator
\begin{align}
T&=\prod_{i} \left(\prod_{j(\neq i)}^{N} (z_i-z_j)^{(\eta-1)}\right)^{n_i},\nonumber\\
&=\prod_{i} \beta_i^{n_i},
\end{align}
where we have defined $\beta_i=\prod_{j(\neq i)}^{N} (z_i-z_j)^{(\eta-1)}$. Then 
\begin{align}
T\Ket{\psi_q^\eta}=\Ket{\psi_q^1[p]_\infty},
\end{align}
and using Eq.\eqref{eqlambdaedge}, 
\begin{align}
T^{-1} \Lambda_i T \Ket{\psi_q^\eta}=0.
\end{align}
Notice first that
\begin{align}
T^{-1}d_j T&=\prod_{l} \beta_l^{-n_l} d_j \prod_{m} \beta_m^{n_m},\nonumber\\
&=\beta_j^{-n_j} d_j \beta_j^{n_j},\nonumber\\
&=\beta_j d_j.
\end{align}
This means that we can rewrite $T^{-1} \Lambda_i T$ as
\begin{align}
\Lambda'_i &\equiv T^{-1} \Lambda_i T,\nonumber\\
&=\sum_{j(\neq i)} \frac{1}{z_i-z_j} [\beta_j d_j- \beta_i d_i (qn_j-1)].
\end{align}
Finally we get that the state $\psi_q^\eta$ is a ground state of the Hamiltonian
\begin{align}
H_q^\eta=\sum_i \Lambda_i^{'\dagger} \Lambda'_i,
\end{align}
and exact diagonalization on small systems shows that it is its only ground state of total number of particle $M=\eta\frac{N}{q}$. Note that in this derivation we assumed $p>-q$ to be able to use the Hamiltonian for the edge state $\psi_q^1[p]_\infty$. This implies that this derivation is valid whenever $N(1-\eta)>-q$, in other words $\eta<1+\frac{q}{N}$. In the thermodynamic limit, the Hamiltonian is therefore valid for $\eta\leq 1$ (see Fig.~\ref{parenthamiltonian}).\\

\subsection{Parent Hamiltonians for the states at $q=4$}

The parent Hamiltonians derived in the previous section are valid for $\eta<1+\frac{q}{N}$, and are therefore not valid for all the values of $\eta$, unless $q=2$. In this section we derive a Hamiltonian valid for all values of $\eta$ at $q=4$. We first concentrate on the half-filled case $q=4$, $\eta=2$.

Using a similar procedure as in Section \ref{parentH1} we derive in Appendix \ref{appendixB} an operator $\Omega_i$ annihilating the wave function $\psi_4^2$:
\begin{equation}
\label{omegaH42}
\Omega_i=\sum_{j(\neq i)}\sum_{k(\neq j)} \frac{1}{z_k-z_j} n_i d_j s_k,
\end{equation}
where $d_j$ is the bosonic annihilation operator at site $j$, $n_i=d_i^\dagger d_i$ and $s_k=2n_k-1$ is the corresponding spin-$1/2$ degree of freedom at site $k$. In the rest of this section we will use both the notations $n_k$ and $s_k$ for brevity.

In addition, the wave-function is particle-hole symmetric and is also annihilated by the particle-hole transformed operator
\begin{equation}
\bar{\Omega}_i=-\sum_{j(\neq i)}\sum_{k(\neq j)} \frac{1}{z_k-z_j} (1-n_i) d_j^\dagger s_k,
\end{equation}
where we have replaced all $d_l$ by $d_l^\dagger$, $d_l^\dagger$ by $d_l$ and $n_l$ by $1-n_l$ (or $s_l$ by $-s_l$).

Finally we define a Hamiltonian
\begin{align}
H_4^2=\sum_i \left[\Omega_i^{\dagger} \Omega_i + \bar{\Omega}_i^{\dagger} \bar{\Omega}_i\right],
\end{align}
which gives
\begin{align}
H_4^2=\sum_i \sum_{j,l (\neq i)} &\sum_{k (\neq j)} \sum_{m (\neq l)} \frac{1}{(z_k^*-z_j^*)(z_m-z_l)}\nonumber\\
& \times s_k \left(n_i d_j^\dagger d_l+(1-n_i)d_j d_l^{\dagger}\right) s_m,
\end{align}
which can be expended and simplified, up to a global irrelevant real factor, as 
\begin{align}
\label{parentH42}
H_4^2=\sum_{j,l} \sum_{\substack{m (\neq l) \\ k (\neq j)}} \frac{1}{(z_k^*-z_j^*)(z_m-z_l)} s_k \left(d_j^\dagger d_l+d_j d_l^{\dagger}\right) s_m,
\end{align}
which is a $4$-body Hamiltonian for the state $\psi_4^2$. Note that since the state $\psi_4^2$ is equivalent to the $SO(2)$ spin state defined in Ref~\onlinecite{Tu2013}, this derivation also provides a parent Hamiltonian for this case which was left open.

Replacing the operators $d_j$ in Eq.\eqref{omegaH42} by $\beta_j d_j$ as in \ref{parentHeta} now leads to operators annihilating the $\psi_4^\eta$ wave function for any value of $\eta$, and thus to parent Hamiltonians for these states. We observe however numerically that this construction does not always give a single ground state of the parent Hamiltonian when $\eta\neq 2$ (see Fig.\ref{parenthamiltonian}). The same construction cannot be applied to other values of $q$ at half-filling, since for $q$ odd there are branch cuts appearing in the null fields such that no operator annihilating the wave function can be derived in this way. For larger even values of $q$, the operator product expansions lead to more terms involving derivative of the fields and these cannot be simplified to involve derivatives of the wave function, as was done here when $q=4$.

\subsection{Conformal transformations of the parent Hamiltonians}

The wave functions defined from correlators of conformal fields are invariant under general M\"{o}bius transformations of the lattice coordinates :
\begin{align}
M:z\rightarrow \frac{az+b}{cz+d}, \ \ \ ad-bc=1.
\end{align}
This is not the case of the Hamiltonians derived previously. In this section we show that there is in fact a class of Hamiltonians annihilating the lattice Laughlin states. These Hamiltonians are related by conformal transformations and some of them have the same symmetries as the lattices considered in this work, but there is no non-zero Hamiltonian that is invariant under all conformal transformations. 

The first observation is that the operators annihilating the wave functions have the form $\Gamma_i=\sum_{j(\neq i)} \lambda_{ij}f_{ij}$, where  $f_{ij}$ does not depend on the coordinates $z_i$ and $\lambda_{ij}=1/(z_i-z_j)$. We show in Appendix \ref{appendixC} that the space of operators annihilating the wave functions that can be obtained by applying M\"{o}bius transformations and multiplying by constant factors the operator $\Gamma_i$ is the space of operators
\begin{align}
\Gamma_{i}^{\alpha\beta\gamma}= \sum_{j(\neq i)} \left(\alpha \lambda_{ij}+\beta \kappa_{ij}+\gamma \rho_{ij}\right) f_{ij},
\end{align}
where $\alpha$, $\beta$ and $\gamma$ are complex numbers and $\kappa_{ij}=\frac{z_i z_j}{z_i-z_j}$, $\rho_{ij}=\frac{z_i+z_j}{z_i-z_j}$. It is also shown that in the class of Hamiltonians obtained from $\Gamma_{i}^{\alpha\beta\gamma}$, there is no non-zero Hamiltonian invariant under all conformal transformations, while the Hamiltonians of the form $\sum_i \sum_{k(\neq i)} \sum_{j(\neq i)} \rho_{kj}^* \rho_{ij}f_{kj}^\dagger f_{ij}$ are invariant under the symmetries of the lattice on the cylinder (Fig.~\ref{fig:cylinder}). This result can be applied to all the parent Hamiltonians derived in this paper to make them invariant under the symmetry transformations of the cylinder.

\section{Conclusion}
\label{SEC:Conclusion}

This work studies lattice versions of Laughlin wave functions with number of particles per flux $\nu=1/q$ for bosons and fermions on arbitrary lattices with filling factor $\mu=\eta/q$, thus allowing to investigate lattice effects all the way along the interpolation between the continuum limit and the completely filled lattice. These wave functions $\psi_q^\eta$ are defined as correlators from CFT fields and can be investigated numerically using Monte-Carlo techniques. 

The phase diagram in the ($q$,$\eta$) plane was established by computing correlations, topological entanglement entropies and braiding properties of localized quasihole excitations. When the lattice is not half-filled, it was shown on the square lattice that the lattice Laughlin states have the same topological properties as the continuum Laughlin state. When the lattice is half-filled, the states have the additional property that they are symmetric under particle-hole transformation. In this case we observed that on the square lattice the states for $q\leq 4$ still share the same topological properties as the continuum Laughlin states, but that the states for $q\geq 5$ are anti-ferromagnets with long-range order. This effect does not persist if the lattice is deformed from a square to a triangular lattice, or on the Kagome lattice, in which case frustration destroys the long-range order and the topological properties of the state are recovered.

We then defined edge states wave functions for these lattice Laughlin states. These edge states are defined by changing the number of particles while keeping the expression of the wave function identical, which is equivalent to introducing a charge operator at infinity in the state. Numerical investigations showed that these edge states have a modified density at the edge, but share the same correlations in the bulk as the lattice Laughlin states.

An exact parent Hamiltonian for which all these edge states are ground states with different number of particles was derived. In addition this Hamiltonian admits the lattice Laughlin state $\psi_q^1$ as its unique ground state of filling factor $1/q$. We then constructed parent Hamiltonians for the lattice Laughlin states at other fillings of the lattice, when $\mu\leq 1/q$ or $\mu\geq (q-1)/q$, which significantly extends the group of lattice FQH models for which parent Hamiltonians can be found. In addition we also derived parents Hamiltonians for the state at $q=4$ for any filling factor of the lattice. We showed more generally that there is a class of Hamiltonians having the lattice Laughlin states as their ground state and that it was possible to find a Hamiltonian in this class that has the symmetries of the lattice on the cylinder, but not a Hamiltonian that is invariant under all conformal transformations of the lattice. 

These parent Hamiltonians have long-range interactions and are difficult to implement in experiments, which raises the question whether there exist local Hamiltonians having the lattice Laughlin states as their ground state. It was already shown that this is the case for the state $\psi_2^1$ in Ref.~\onlinecite{Nielsen2013} and a numerical procedure to deform long-range parent Hamiltonians such as the ones derived in this work was given in Ref.~\onlinecite{Glasser2015}. We note that one candidate for realizing the $\psi_2^{2/3}$ state would be the Hamiltonian investigated in Ref.~\onlinecite{Zhu2015} on the Kagome lattice. 

It is in general not clear how a lattice may affect the properties of Fractional Quantum Hall wave functions. In the present work we provided a model that allows us to consider all lattice filling factors from the continuum limit to the almost filled lattice and to check that the properties of the Laughlin states remain the same except for very special cases. This procedure can also be applied to other Fractional Quantum Hall states that are constructed from wave functions on lattices. It is interesting because it shows that such Fractional Quantum Hall states could possibly be realized in settings where lattice effects are strong and the number of particles per lattice much larger than $\nu=1/q$, especially on frustrated lattices, thus increasing the chance to find new models displaying Fractional Quantum Hall physics. In addition, we showed that by looking at the screening properties of a quasihole wave function, we could probe the topology of the ground state wave function and construct the phase diagram. This method can therefore be used as a tool to probe topology in cases where an ansatz for the quasiparticle excitations is known.

\begin{acknowledgments}
The authors would like to thank Iv\'an Rodr\'iguez, Nicolas Regnault and Benedikt Herwerth for discussions.
This work was supported by the EU integrated project SIQS, the Villum Foundation, FIS2012-33642, QUITEMAD+ (CAM), the Severo Ochoa Program.
\end{acknowledgments}

\allowdisplaybreaks
\appendix
\section{Operators annihilating the $\psi_q^1[p]_\infty$ wave functions}
\label{appendixA}

In this section we obtain operators annihilating the $\psi_q^1[p]_\infty$ wave functions by extending the procedure of Ref.~\onlinecite{Nielsen2015} to $p>-q$, $p$ integer.
It was shown in Ref.~\onlinecite{Tu2014} that 
\begin{align}
\chi(w) &=\oint_{w}\frac{dz}{2\pi i}\frac{1}{z-w} [G^{+}(z)V_{-}(w) - qJ(z)V_{+}(w)],
\end{align}
is a null field, where $G^{+}(z)=:e^{i\sqrt{q}\phi (z)}:$, $J(z)=i\partial_z \phi /\sqrt{q}$, $V_{-}(z)=:e^{-i\frac{1}{\sqrt{q}}\phi (z)}:$ and $V_{+}(z)=:e^{i\frac{q-1}{\sqrt{q}}\phi (z)}:$. The correlator with this field inserted vanishes :
\begin{align}
\Braket{W_{w}(\infty)V_{n_1}(z_1) \dots V_{n_{i-1}}(z_{i-1}) \chi(z_i)\nonumber\\ \times V_{n_{i+1}}(z_{i+1}) \dots V_{n_N}(z_N)}=0.
\end{align}
This expression is the sum of two terms,
\begin{align}
\oint_{z_i}\frac{dz}{2\pi i}\frac{1}{z-z_i}  &\Braket{W_{p}(w)V_{n_1}(z_1) \dots V_{n_{i-1}}(z_{i-1})\nonumber\\ \times & G^{+}(z)V_{-}(z_i) V_{n_{i+1}}(z_{i+1}) \dots V_{n_N}(z_N)},
\end{align}
and
\begin{align}
-q\oint_{z_i}\frac{dz}{2\pi i}\frac{1}{z-z_i} & \Braket{W_{p}(w)V_{n_1}(z_1) \dots V_{n_{i-1}}(z_{i-1})\nonumber\\ \times & J(z)V_{+}(z_i) V_{n_{i+1}}(z_{i+1}) \dots V_{n_N}(z_N)}.
\end{align}
The second term, multiplied by $(-1)^{i-1} \Ket{n_1,\ldots,n_{i-1},0,n_{i+1},\ldots,n_N}$ and after summation of all $n_k$, gives\cite{Nielsen2015} 
\begin{align}
\label{eqappendix1}
 -\left(\sum_{j(\neq i)} \frac{1}{z_i-z_j} d_i (qn_j-1) + \frac{p}{z_i-w} d_i\right)\Ket{\psi_q^1[p(w)]},
\end{align}
where $d_j$ is the hardcore boson (resp. fermion) annihilation operator for $q$ even (resp. odd), and $n_j=d_j^\dagger d_j$ is the number of particles at site $j$.
The first term instead becomes, after deforming the contour integral around each $z_j$,
\begin{align}
-\oint_{w}&\frac{dz}{2\pi i}\frac{1}{z-z_i}  \Braket{W_{p}(w)V_{n_1}(z_1) \dots V_{n_{i-1}}(z_{i-1})\nonumber\\  & \qquad \qquad \times G^{+}(z)V_{-}(z_i) V_{n_{i+1}}(z_{i+1}) \dots V_{n_N}(z_N)}\nonumber\\
&-\sum_{j\neq i} \oint_{z_j}\frac{dz}{2\pi i}\frac{1}{z-z_i}  \Braket{W_{p}(w)V_{n_1}(z_1) \dots V_{n_{i-1}}(z_{i-1})\nonumber\\  & \qquad \qquad \times G^{+}(z)V_{-}(z_i) V_{n_{i+1}}(z_{i+1}) \dots V_{n_N}(z_N)}.\nonumber\\
\end{align}
The sum in the second term has been computed in Ref.~\onlinecite{Nielsen2015} and gives, after multiplication by $(-1)^{i-1} \Ket{n_1,\ldots,0,\ldots,n_N}$,
\begin{align}
\label{eqappendix2}
\sum_{j(\neq i)} \frac{1}{z_i-z_j} d_j \Ket{\psi_q^1[p(w)]}.
\end{align}
The last remaining term is 
\begin{align}
-&(-1)^{i-1+p}\sum_{n_i} \delta_{n_i=0} \oint_{w}\frac{dz}{2\pi i}\frac{(-1)^{q\sum_{k=1}^{i-1}n_k}}{z-z_i} \nonumber\\
&\times \Braket{G^{+}(z) W_{p}(w)V_{n_1}(z_1) \dots V_{n_N}(z_N)},
\end{align}
where we have commuted $G^{+}(z)$ in front of the correlator.
Since we know the expression of the correlator, we can compute the contour integral. The expression then simplifies to
\begin{align}
-&(-1)^{i-1+p} \delta_{p<0} \sum_{n_i} n_i \frac{1}{(-p-1)!}(w-z_i)^{-p} 
\nonumber\\
&\times \prod_{k(\neq i)} (z_i-z_k)^{-(qn_k-1)} \nonumber\\
&\times\lim\limits_{z\rightarrow w} \frac{\text{d}^{-p-1}}{\text{d}z^{-p-1}}\prod_j (z-z_j)^{qn_j-1-(q+1)\delta_{ij}}\nonumber\\
&\times \Braket{W_{p}(w)V_{n_1}(z_1) \dots V_{n_N}(z_N)}.
\end{align}
Now observe that this expression is zero when $p>0$ (due to the first delta factor), but also when $p>-q$ (due to the derivative and the exponent of the polynomial). This shows that when $p>-q$, this term does not contribute and the resulting expression is given by summing Eq.~\eqref{eqappendix1} and Eq.~\eqref{eqappendix2}, which leads to 
\begin{align}
\Lambda_i \Ket{\psi_q^1[p(w)]}=0,
\end{align}
where \begin{align}
\Lambda_i=\sum_{j(\neq i)} \frac{1}{z_i-z_j} [d_j-d_i (qn_j-1)] - \frac{p}{z_i-w} d_i.
\end{align}

\section{Operators annihilating the $\psi_4^2$ wave function}
\label{appendixB}

In this section we obtain operators annihilating the $\psi_4^2$ wave function.
Using the fields defined in Section \ref{parentH1}, we have for $q=4$ and $\eta=2$ the following operator product expansions :
\begin{align}
G^{+ }(z)V_{-}(w)&= \frac{1}{(z-w)^2} e^{i(2\phi (z)-\phi (w))},\nonumber\\
&\sim\frac{1}{(z-w)^2} e^{i\phi (w)}+\frac{1}{(z-w)}2i \partial \phi (w) e^{i\phi (w)}\nonumber\\
&\sim\frac{1}{(z-w)^2} V_{+}(w)+\frac{1}{(z-w)}2 \partial V_{+}(w),
\end{align}
so that
\begin{align}
\label{OPE}
G^{+ }(z)V_{n_{j}=0}(w)&\sim \frac{1}{(z-w)^2} V_{n_{j}=1}(w)\nonumber\\
&\quad +\frac{1}{(z-w)}2 \partial V_{n_{j}=1}(w),\\
G^{+ }(z)V_{n_{j}=1}(w)&\sim 0.
\end{align}
The field 
\begin{align}
\chi _{2}(w)=\oint_{w}\frac{dz}{2\pi i} G^{+}(z)V_{+}(w)
\end{align}
is therefore a null field. We can now use the fact that the correlator with the field inserted vanishes :
\begin{align}
\Braket{V_{n_1}(z_1) &\dots V_{n_{i-1}}(z_{i-1}) \chi(z_i)\nonumber\\ &\times V_{n_{i+1}}(z_{i+1}) \dots V_{n_N}(z_N)}=0.
\end{align}
This can be rewritten as
\begin{align}
&\oint_{z_i}\frac{dz}{2\pi i} 
\langle V_{n_{1}}(z_{1})\ldots G^{+}(z)V_{+}(z_i) \ldots V_{n_{N}}(z_{N})\rangle\nonumber\\
&=\oint_{\infty}\frac{dz}{2\pi i}
\langle V_{n_{1}}(z_{1})\ldots G^{+}(z)V_{+}(z_i) \ldots V_{n_{N}}(z_{N})\rangle\nonumber\\
&\quad-\sum_{j(\neq i)}\oint_{z_j}\frac{dz}{2\pi i}
\langle V_{n_{1}}(z_{1})\ldots G^{+}(z)V_{+}(z_i) \ldots V_{n_{N}}(z_{N})\rangle,
\end{align}
where we have deformed the contour integral around the positions $z_j$. The contour integral at infinity can be computed by evaluating the correlator, which leads to a zero contribution. The remaining part can be transformed by observing that the operators $G^{+}(z)$ and $V_{\pm}(z_j)$ commute, so that
\begin{align}
&-\sum_{j(\neq i)}\oint_{z_j}\frac{dz}{2\pi i}
\langle V_{n_{1}}(z_{1})\ldots G^{+}(z)V_{+}(z_i) \ldots V_{n_{N}}(z_{N})\rangle\nonumber \\
&=-\sum_{j(\neq i)}\oint_{z_j}\frac{dz}{2\pi i}
\langle V_{n_{1}}(z_{1})\ldots G^{+}(z)V_{n_j}(z_j)\ldots\nonumber\\
&\qquad \qquad \qquad \qquad \qquad \qquad \times V_{+}(z_i) \ldots V_{n_{N}}(z_{N})\rangle\nonumber \\
&=-\sum_{j(\neq i)}\oint_{z_j}\frac{dz}{2\pi i} \frac{\sum_{n'_j}d_{n_jn'_j}}{(z-z_j)^2}
\langle V_{n_{1}}(z_{1})\ldots V_{n'_j}(z_j) \ldots\nonumber\\
&\qquad \qquad \qquad \qquad \qquad \qquad \times V_{+}(z_i) \ldots V_{n_{N}}(z_{N})\rangle\nonumber\\
&\phantom{=}-\sum_{j(\neq i)}\oint_{z_j}\frac{dz}{2\pi i} 2 \frac{\sum_{n'_j}d_{n_jn'_j}}{z-z_j}
\langle V_{n_{1}}(z_{1})\ldots \partial V_{n'_j}(z_j) \ldots\nonumber\\
&\qquad \qquad \qquad \qquad \qquad \qquad \times  V_{+}(z_i) \ldots V_{n_{N}}(z_{N})\rangle,
\end{align}
where we have moved the operator $G^{+}(z)$ to position $j$ and applied the operator product expansion in Eq.~\eqref{OPE}. $d_{n_jn'_j}$ are the matrix elements of the bosonic annihilation operator $d_j$ at site $j$ ($d_{00}=d_{11}=d_{10}=0$, $d_{01}=1$), so that only terms having a non-zero operator product expansion contribute to this expression. The contour integral evaluated in the first term gives zero, so that we are left with
\begin{align}
&-2 \sum_{j(\neq i)}\sum_{n'_j}d_{n_jn'_j}
\langle V_{n_{1}}(z_{1})\ldots \partial V_{n'_j}(z_j) \ldots\nonumber\\
&\qquad \qquad \qquad \qquad \times   V_{+}(z_i) \ldots V_{n_{N}}(z_{N})\rangle\nonumber\\
&=-2 \sum_{j(\neq i)} \sum_{n'_j}d_{n_jn'_j}
\frac{\partial}{\partial z_j} \langle V_{n_{1}}(z_{1})\ldots\nonumber\\
&\qquad \qquad \qquad \qquad \times   V_{n'_j}(z_j) \ldots  V_{+}(z_i) \ldots V_{n_{N}}(z_{N})\rangle\nonumber\\
&=-2 \sum_{j(\neq i)} \sum_{n'_j}d_{n_jn'_j}
\frac{\partial}{\partial z_j} \psi_4^2(n_{1},\ldots,n'_j,\dots\nonumber\\
&\qquad \qquad \qquad \qquad \qquad \qquad \dots,n_i=1,\ldots,n_N),
\end{align}
where we have used the expression of the wave function $\psi_4^2$ as a correlator of fields. Let us now compute the derivative of the wave function :
\begin{align}
\frac{\partial}{\partial z_j}& \psi_4^2(n_{1},\ldots,n'_j,\dots,n_i=1,\ldots,n_N)\nonumber\\
&=\psi(n_{1},\ldots,n_N)\frac{\partial}{\partial z_j} \ln(\psi_4^2(n_1,\ldots,n_N))\nonumber\\
&=\psi(n_{1},\ldots,n_N)\frac{\partial}{\partial z_j}\left[\sum_{k(<j)} s_k s'_j \ln(z_k-z_j)\right.\nonumber\\
&\qquad \qquad \qquad \left.+\sum_{k(>j)} s_k s'_j \ln(z_j-z_k)\right]\nonumber\\
&=\psi(n_{1},\ldots,n_N)\left[\sum_{k(\neq j)} s_k s'_j \frac{-1}{z_k-z_j}\right],
\end{align}
where $s_k=2n_k-1$, $s'_j=2n'_j-1$  are the corresponding spin-$1/2$ degree of freedom at site $k$ and $j$. In the rest of this section we will use both the notations $n_k$ and $s_k$ for brevity. Note that $k$ can be equal to $i$ in this sum, in which case $s_i=1$ since $n_i=1$. The previous expression becomes
\begin{align}
2 &\sum_{j(\neq i)} \sum_{n'_j}d_{n_jn'_j} \sum_{k(\neq j)} s_k s'_j \frac{1}{z_k-z_j} \psi_4^2(n_{1},\ldots,n'_j,\ldots\nonumber\\
&\qquad \qquad \qquad \qquad \qquad \dots,n_i=1,\ldots,n_N)\rangle.
\end{align}
Since this expression is zero unless $n_j=0$ and $n'_j=1$, we can replace $s'_j$ by 1. To take into account the fact that this expression evaluates to $0$ when $n_i=0$, we introduce the number operator $n_i$ and write the expression as
\begin{align}
2 &\sum_{j(\neq i)} \sum_{n'_j}d_{n_jn'_j} n_i \sum_{k(\neq j)} s_k \frac{1}{z_k-z_j}\psi_4^2(n_{1},\ldots,n'_j,\ldots\nonumber\\
&\qquad \qquad \qquad \qquad \qquad \dots,n_i,\ldots,n_N)\rangle,
\end{align}
which when multiplied by the basis elements $|n_1,\ldots,n_N\rangle$ and summed over all $n_k$ leads to 
\begin{equation}
\sum_{j(\neq i)}\sum_{k(\neq j)} \frac{1}{z_k-z_j} n_i d_j s_k |\psi_4^2\rangle.
\end{equation}
We started with the fact that this expression was zero, so the operator 
\begin{equation}
\Omega_i=\sum_{j(\neq i)}\sum_{k(\neq j)} \frac{1}{z_k-z_j} n_i d_j s_k
\end{equation}
annihilates the wave function $\psi_4^2$.

\section{Conformal transformations of the parent Hamiltonians}
\label{appendixC}
In this section we discuss the properties of the parent Hamiltonians obtained in this work under conformal transformations of the coordinates $z_i$. We will use the notations :
\begin{align}
\lambda_{ij}=\frac{1}{z_i-z_j},\ \ \
\kappa_{ij}=\frac{z_i z_j}{z_i-z_j},\ \ \
\rho_{ij}=\frac{z_i+z_j}{z_i-z_j}.
\end{align}
If an operator $\Gamma_i$ has the form $\Gamma_i=\sum_{j(\neq i)} \lambda_{ij}f_{ij}$, where $f_{ij}$ does not depend on the coordinates $z_i$, then it is transformed under a M\"{o}bius transformation $M$ as 
\begin{align}
\Gamma_{i}\rightarrow \sum_{j(\neq i)} \left(d^2 \lambda_{ij}+c^2 \kappa_{ij}+cd \rho_{ij}\right) f_{ij}.
\end{align}
If $\Gamma_i$ annihilates a wave function that is invariant under $M$, then $\sum_{j(\neq i)} \kappa_{ij}f_{ij}$ annihilates the same wave function (choice $c=1$, $d=0$ in the previous equation). Since $\rho_{ij}=z_i \lambda_{ij} + \frac{1}{z_i} \kappa_{ij}$, then $\sum_{j(\neq i)} \rho_{ij}f_{ij}$ also annihilates the wave function. Noting that the composition of M\"{o}bius transformations is a M\"{o}bius transformation, the space of operators annihilating the wave function that can be obtained by applying M\"{o}bius transformations and multiplying by constant factors the operator $\Gamma_i$ is therefore the space of operators
\begin{align}
\Gamma_{i}^{\alpha\beta\gamma}= \sum_{j(\neq i)} \left(\alpha \lambda_{ij}+\beta \kappa_{ij}+\gamma \rho_{ij}\right) f_{ij},
\end{align}
where $\alpha$, $\beta$ and $\gamma$ are complex numbers. An operator in this space is invariant under $M$ when the following conditions are satisfied :
\begin{align}
\alpha&= d^2 \alpha + b^2 \beta + 2bd \gamma,\nonumber\\
\beta&= c^2 \alpha + a^2 \beta + 2ac \gamma,\\
\gamma&= cd \alpha + ab \beta + (ad+bc) \gamma\nonumber.
\end{align}
Note that since the Hamiltonians we construct have the form $\sum_i \Gamma_i^\dagger \Gamma_i$, these conditions only need to be satisfied up to a phase for the Hamiltonian to be invariant under $M$.
A particular case is the translation $z\rightarrow z+1$ ($a=1, b=1, c=0, d=1$), which leads to the conditions $\beta=0$, $\gamma=0$. Another particular case are the rotations along the periodic direction of the cylinder $z\rightarrow e^{2\pi i/N_y}z$, ($a=e^{2\pi i/N_y}, b=0, c=0, d=1$), which lead to the conditions $\alpha=0$, $\beta=0$. Therefore $\Gamma_i$ cannot be invariant under translations and rotations at the same time (unless $\Gamma_i=0$), hence it cannot be invariant (up to a phase) under all conformal transformations. We note moreover that for the operators $f_{ij}$ that we have used before (in particular the case $f_{ij}=d_j-d_i (qn_j-1)$) no further simplification in $H=\sum_i \Gamma_i^\dagger \Gamma_i$ appears, so that in the space of Hamiltonians we constructed there is no non-zero Hamiltonian that would be invariant under all conformal transformations.

However the choice $\alpha=0$, $\beta=0$ leads to an operator that is invariant under rotations along the periodic direction of the cylinder. In addition the operator $\Gamma_i=\sum_{j(\neq i)} \rho_{ij}f_{ij}$ is then invariant, up to a phase, under a transformation $z\rightarrow 1/z$. This means that the Hamiltonians of the form $\sum_i \sum_{k(\neq i)} \sum_{j(\neq i)} \rho_{kj}^* \rho_{ij}f_{kj}^\dagger f_{ij}$ are invariant under the symmetries of the lattice on the cylinder (Fig.~\ref{fig:cylinder}).
\bibliography{biblio}

\end{document}